\documentclass[11pt]{article}
\usepackage{graphicx}
\usepackage{epsfig}
\usepackage{graphicx,colordvi}
\usepackage[dvips]{color}
\usepackage{amssymb}
\usepackage{amsmath,amsfonts}
\usepackage{url}
\usepackage{bm}
\usepackage{longtable}

\newcommand{\hmo}{HoMnO$_3\:$}

\begin{document}
\parskip 1ex


\vspace*{2cm}

\rule{16.0cm}{1mm}
\vspace{2mm}
\setcounter{section}{0}
\setcounter{figure}{0}

\begin{center}

{\bf \Large First Principles Studies of Multiferroic Materials}

Silvia Picozzi$^1$ and Claude Ederer$^2$

$^1$Consiglio Nazionale delle Ricerche - Istituto Nazionale per la
Fisica della Materia (CNR-INFM), CASTI Regional Laboratory, 67100
L'Aquila, Italy \\[0.5mm]
$^2$School of Physics, Trinity College, Dublin 2, Ireland

\end{center}

\begin{abstract}

Multiferroics, materials where spontaneous long-range magnetic and
dipolar orders coexist, represent an attractive class of compounds,
which combine rich and fascinating fundamental physics with a
technologically appealing potential for applications in the general
area of spintronics. {\em Ab-initio} calculations have significantly
contributed to recent progress in this area, by elucidating different
mechanisms for multiferroicity and providing essential information on
various compounds where these effects are manifestly at play. In
particular, here we present examples of density-functional theory
investigations for two main classes of materials: a) {\em proper}
multiferroics (where ferroelectricity is driven by hybridization or
purely structural effects), with BiFeO$_3$ as prototype material, and
b) {\em improper} multiferroics (where ferroelectricity is driven by
correlation effects and is strongly linked to electronic degrees of
freedom such as spin, charge, or orbital ordering), with rare-earth
manganites as prototypes. As for proper multiferroics,
first-principles calculations are shown to provide an accurate
qualitative and quantitative description of the physics in BiFeO$_3$,
ranging from the prediction of large ferroelectric polarization and
weak ferromagnetism, over the effect of epitaxial strain, to the
identification of possible scenarios for coupling between
ferroelectric and magnetic order. For the class of improper
multiferroics, {\em ab-initio} calculations have shown that, in those
cases where spin-ordering breaks inversion symmetry ({\it i.e.} in
antiferromagnetic E-type HoMnO$_3$), the magnetically-induced
ferroelectric polarization can be as large as a few $\mu$C/cm$^2$. The
presented examples point the way to several possible avenues for
future research: On the technological side, first-principles
simulations can contribute to a {\em rational materials design}, aimed
at identifying spintronic materials that exhibit ferromagnetism and
ferroelectricity at or above room-temperature. On the fundamental
side, {\em ab-initio} approaches can be used to explore new mechanisms
for ferroelectricity by exploiting electronic correlations that are at
play in transition metal oxides, and by suggesting ways to maximize
the strength of these effects as well as the corresponding ordering
temperatures.

\end{abstract}

\section{Introduction to  multiferroic materials}

Recent years have seen an enormous increase in research activity in
the field of multiferroic materials and magneto-electric effects. In
December 2007 Science Magazine listed multiferroic materials as one
out of ten ``Areas to watch in 2008'', the only entry from the
Materials Science/Condensed Matter area that was included in this
list. First principles calculations using density functional theory
(DFT) \cite{Hohenberg/Kohn:1964,Kohn/Sham:1965,Martin:Book} have
played an important role in this ``Renaissance of Magnetoelectric
Multiferroics'' \cite{Spaldin/Fiebig:2005}. In the present paper we
give a brief summary of the current status of research on multiferroic
materials and highlight some of the contributions that have been made
using first principles electronic structure calculations.

According to the original definition put forward by Hans Schmid
\cite{Schmid:1994}, multiferroic materials are materials that combine
two or more of the primary forms of ferroic order, { \it i.e.}
ferroeleasticity, ferroelectricity, ferromagnetism, and
ferrotoroidicity. In practice, most of the recent research has focused
on materials that combine some form of magnetic order (ferromagnetic,
antiferromagnetic, non-collinear, \dots) with
ferroelectricity. Therefore, the term \emph{multiferroics} is nowadays
often used synonymous with \emph{magnetic ferroelectrics}.

Research on multiferroics (or magnetic ferroelectrics) is also
intimately interwoven with research on the \emph{magneto-electric
  effect}, which is the property that in certain materials a magnetic
field induces an electric polarization and, conversely, an electric
field induces a magnetization. Traditionally, one distinguishes
between linear, quadratic, and higher order magneto-electric effects
\cite{Schmid:1973}, but more recently the term ``magneto-electric
effect'' is often (mis-)used to describe any form of cross-correlation
between magnetic and (di-)electric properties. (For example, when the
application of an external magnetic field induces a phase transition
between ferroelectric/non-ferroelectric phases.) It is important to
point out, though, that not every magnetic ferroelectric exhibits a
linear magneto-electric effect (in the original sense) and that not
every material that exhibits a linear magneto-electric effect is also
simultaneously multiferroic.

Due to the combination of magnetic and dielectric properties, with
eventual cross-coupling between these properties, multiferroics have
immense potential for technological device applications and at the
same time they pose very interesting and rich fundamental physics
problems. It is probably this combination of applied and fundamental
research that is partly responsible for the strong attraction that
these materials have developed in recent years.

Multiferroics form a very diverse class of materials, and there is no
unique ``theory of multiferroics''. Nearly every material has to be
studied on its own right, and eventually involves very different
physical mechanisms than other multiferroic materials. However, it has
proven to be very useful to classify different multiferroics according
to the mechanism that drives the ferroelectricity in the corresponding
systems. In particular two major classes of multiferroics can be
distinguished:
\begin{enumerate}
\item{Multiferroics, where the ferroelectricity is driven by
  hybridization and covalency or other purely structural effects.}
\item{Multiferroics, where the ferroelectricity is driven by some
  other electronic mechanism, e.g. ``correlation'' effects.}
\end{enumerate}
In the second case, ferroelectricity always arises as a secondary
effect that is coupled to some other form of ordering, such as
magnetic or charge ordering. Therefore, these systems are often called
``improper magnetic ferroelectrics''. We note that also in the first
class at least one material, hexagonal YMnO$_3$, has been classified
as an improper ferroelectric, where the electric polarization is not
the primary order parameter, but instead is coupled to a different
non-polar structural instability \cite{Fennie/Rabe_YMO:2005}. In spite
of that, and for the purpose of this article, we will call materials
belonging to the first category ``proper magnetic ferroelectrics'' (or
``proper multiferroics''), whereas materials in the second category
will be called ``improper magnetic ferroelectrics'' (or ``improper
multiferroics''). We note that this ``zoology'' of multiferroics is
still work in progress, and that the discovery of new materials might
require a further refinement or redefinition of previous
classifications.

In this article, we are not attempting to provide a complete review of
all first principles work that has been carried out so far. Instead,
we discuss some specific examples that illustrate the power of these
methods in elucidating the physical origins of the observed properties
of known multiferroics, and point out the possibilities in predicting
novel effects and designing new materials with optimized
properties. Also, we focus only on single-phase (bulk) materials,
therefore leaving out all those effects coming from the combination of
ferroelectrics and ferromagnets in (artificial) multiferroic
heterostructures.  Several excellent review articles about general
aspects of multiferroic materials and magneto-electric effects have
already been published, see for example
Refs.~\cite{Smolenskii/Chupis:1982,Fiebig:2005,Eerenstein/Mathur/Scott:2006,Khomskii:2006,Cheong/Mostovoy:2007,Ramesh/Spaldin:2007},
and much of the early first principles work has also been reviewed in
Ref.~\cite{Hill:2002}, and more recently in
Ref.~\cite{Ederer/Spaldin_COSSMS:2006}.

The remainder of this article is structured as follows: we start by
giving a more detailed discussion of proper magnetic ferroelectrics,
and we summarize some of the key developments where first principles
studies have made important contributions. We then focus in particular
on research related to BiFeO$_3$, which is probably the most studied
multiferroic material to date. After that, we give an overview over
more recent advancements in the field of improper multiferroics, and
discuss some recent work on various manganite systems: orthorhombic
E-type HoMnO$_3$ and half-doped La$_{0.5}$Ca$_{0.5}$MnO$_3$. We end
with some conclusions and perspectives for future research.

Finally, before starting our discussion of proper and improper
multiferroics, we want to mention that even though no new
calculational techniques have to be developed for the study of these
materials, research on multiferroics typically involves a combination
of a variety of advanced techniques, most of which have been
established only during the last decade (roughly speaking). These
techniques include for example beyond-LDA/GGA approaches for the
treatment of strongly correlated transition metal oxides, mostly
LSDA+$U$
\cite{Anisimov/Zaanen/Andersen:1991,Anisimov/Aryatesiawan/Liechtenstein:1997},
methods for the treatment of non-collinear magnetism \cite{nonco} and
spin-orbit coupling \cite{soc}, the Berry phase approach to calculate
electric polarization \cite{King-Smith/Vanderbilt:1993,Resta:1994}
combined with a further analysis using maximally localized Wannier
functions \cite{Marzari/Vanderbilt:1997}, and many more.

\section{Proper magnetic ferroelectrics}

Most of the ``early'' first principles work on multiferroics was
focused on proper magnetic ferroelectrics, in particular on
identifying mechanisms for ferroelectricity that are compatible with
the simultaneous presence of magnetic order.

In conventional ferroelectrics such as BaTiO$_3$ and PbTiO$_3$,
hybridization effects between the filled oxygen $p$ states and the
empty transition metal $d$ states are essential for the appearance of
the structural instability that causes ferroelectricity
\cite{Cohen:1992}. Early first principles work pointed out that such a
mechanism is unfavorable if the transition metal $d$ states are
partially filled, which to some extent explains the relative scarcity
of magnetic ferroelectrics \cite{Hill:2000,Filippetti/Hill:2002}.

The ferroelectricity in multiferroic materials is therefore generally
caused by a different mechanism than in prototypical ferroelectric
materials such as BaTiO$_3$, PbTiO$_3$, or KNbO$_3$, which all contain
transition metal cations with a formal $d^0$ configuration. As in the
case of these conventional ferroelectrics, electronic structure
calculations have been crucial in identifying and classifying
different mechanisms for ferroelectricity that are also compatible
with the simultaneous presence of partially filled $d$ of $f$ states.

Two such mechanisms have emerged from these early studies:
\begin{enumerate}
\item{Ferroelectricity caused by stereo-chemically active
  ``lone-pair'' cations, e.g. Bi$^{3+}$ or Pb$^{2+}$.}
\item{``Geometric ferroelectricity'', where the structural instability
  is driven by size effects and other geometrical considerations.}
\end{enumerate}

It is well known in chemistry, that cations containing a highly
polarizable 5$s$ or 6$s$ lone pair of valence electrons have a strong
tendency to break local inversion symmetry. This can be understood by
a mixing between $ns$ and $np$ electron states, which can lower the
energy of the cation, but is only allowed if the ionic site is not an
inversion center. Alternatively, in a solid this tendency can be
understood as cross-gap hybridization between occupied oxygen $p$ and
unoccupied $np$ states of the lone-pair cation, similar to the
cross-gap hybridization between occupied oxygen $p$ and unoccupied
transition metal $d$ states that gives rise to the ferroelectricity in
conventional ferroelectrics \cite{Ghita_et_al:2005}. In fact, the
presence of the lone-pair active Pb$^{2+}$ cation is an important
factor for the ferroelectric properties of PbTiO$_3$ (in addition to
the presence of the $d^0$ Ti$^{4+}$ cation) \cite{Cohen:1992}. The
lone-pair mechanism was identified as the source of the ferroelectric
instability in BiMnO$_3$ \cite{Hill/Rabe:1999,Seshadri/Hill:2001} and
BiFeO$_3$ \cite{Neaton_et_al:2005,Ravindran_et_al:2006}.

\begin{figure}
\centerline{\includegraphics[width=0.9\textwidth]{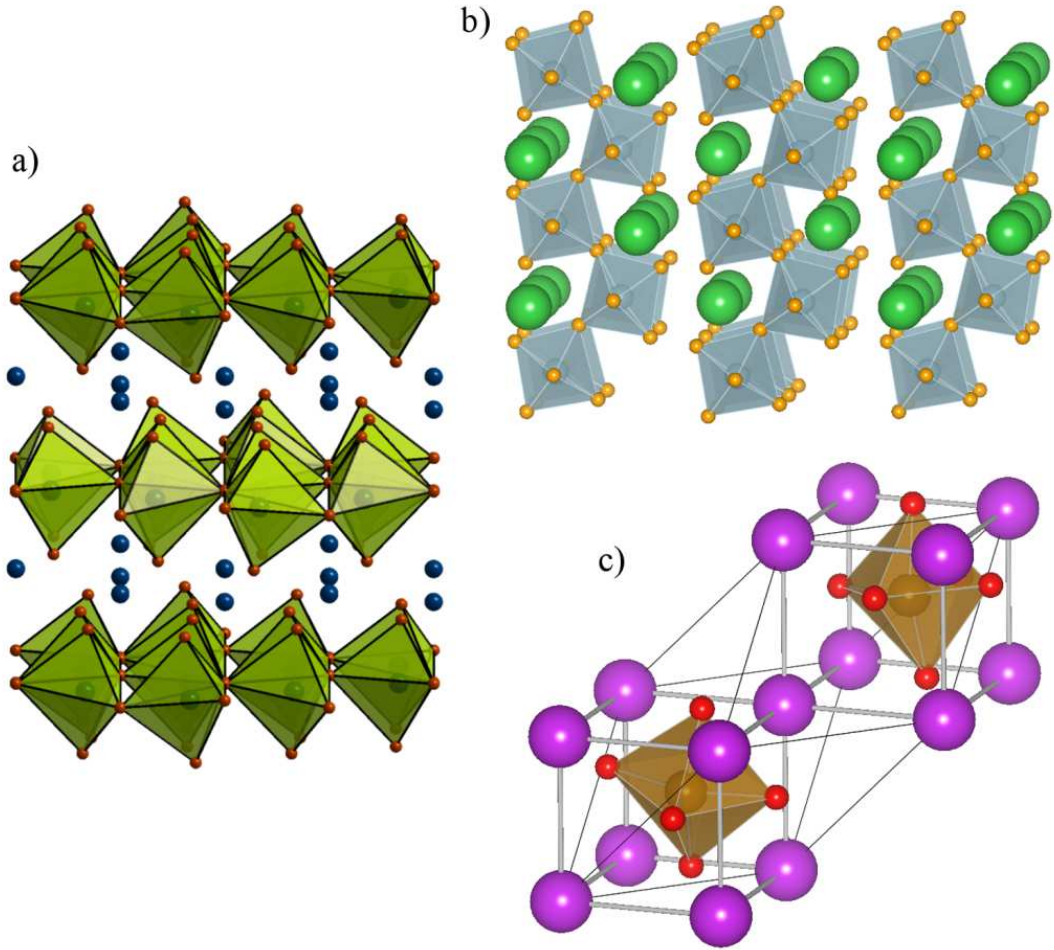}}
\caption{Crystal structures of various magnetic ferroelectrics: a)
  YMnO$_3$, which has been classified as improper geometric
  ferroelectric, crystallizes in a layered hexagonal structure,
  consisting of a two-dimensional arrangement of connected oxygen
  bi-pyramids surrounding the Mn$^{3+}$ cations that are separated by
  layers of Y$^{3+}$ cations. b) BaNiF$_4$, a proper geometric
  ferroelectric, is found in an orthorhombic structure with buckled
  planes of fluorine octahedra around the Ni$^{2+}$ cations and
  additional interjacent Ba$^{2+}$ cations.  c) BiFeO$_3$, where the
  ferroelectricity is driven by the stereochemically-active Bi$^{3+}$
  cation, exhibits a rhombohedrally distorted perovskite structure,
  where all ionic sublattices are displaced relative to each other
  along the polar (111) direction, and the oxygen octahedra are
  rotated around the same (111) axis, alternately clockwise and
  counter-clockwise.}
\label{fig:structures}
\end{figure}

In contrast to this, the ferroelectric instability in geometric
ferroelectrics does not involve any significant re-hybridization
effects. Instead, a structural instability in such systems is
generated mainly by size effects and geometric constraints, {\it i.e.}
the space-filling and ionic coordination in the ``ideal''
high-symmetry structure is not optimal, but can be improved by a small
distortion that eventually breaks inversion symmetry. The first
material that was identified as geometric ferroelectric is hexagonal
YMnO$_3$ \cite{VanAken_et_al:2004} (see
Fig.~\ref{fig:structures}a). First principles calculations showed that
the ferroelectric structure of this material results from an interplay
between a polar $\Gamma$-point mode and a non-polar Brillouin
zone-boundary mode that leads to a unit cell tripling
\cite{VanAken_et_al:2004,Fennie/Rabe_YMO:2005}. Furthermore,
calculated phonon frequencies together with group theoretical analysis
suggests that YMnO$_3$ is an \emph{improper} ferroelectric, where the
hexagonal point group of the centrosymmetric high-symmetry structure
allows a coupling between the otherwise stable $\Gamma_2^-$ and the
unstable $K_3$ mode \cite{Fennie/Rabe_YMO:2005}.

An example for \emph{proper} geometric ferroelectricity has been found
in the series of antiferromagnetic (AFM) fluorides Ba$M$F$_4$, where
$M$ can be Mn, Fe, Co, or Ni \cite{Ederer/Spaldin_BMF:2006}. The
special connectivity of the fluorine octahedra in these systems, which
are arranged in quasi-two-dimensional sheets, gives rise to one
unstable phonon mode that involves alternating octahedral rotations
together with an overall shift of the interjacent Ba cations relative
to the other ions (see Fig.~\ref{fig:structures}b). This shift creates
an electric dipole moment, and since only one structural mode is
involved the corresponding ferroelectricity is classified as
``proper''. Due to the fact that fluorine systems are generally much
more ionic and less covalent than oxides, geometric ferroelectricity
can be expected to be the dominant source for ferroelectric
instabilities in fluoride compounds.

Very recently, the question of why exactly the standard $p$-$d$
hybridization mechanism for ferroelectricity is unfavorable for
systems with partially filled $d$ shells has been revisited
\cite{Bhattacharjee/Bousquet/Ghosez:2009,Rondinelli/Eidelson/Spaldin:arXiv}. For
perovskite systems, with dominantly cubic crystal field splitting
between the $t_{2g}$ and $e_g$ manifolds, it is not fully clear why
for example a $d^3$ configuration with partially filled $t_{2g}$
states, but empty $e_g$ orbitals, cannot give rise to a favorable
cross-gap hybridization between filled oxygen $p$ and empty transition
metal $e_g$ states. It was suggested that the Hund's coupling between
$t_{2g}$ and $e_g$ states will disfavor such hybridization
\cite{Khomskii:2006}. This was supported by LDA+$U$ calculations for
CaMnO$_3$, where the Hund's coupling was effectively ``turned off'',
which indeed resulted in a tendency for off-centering of the Mn$^{4+}$
cation. In addition, recent first principles calculations for
CaMnO$_3$, SrMnO$_3$, and BaMnO$_3$ in the perovskite structure show
that these systems can develop a ferroelectric instability, but that
this ferroelectric instability competes with a non-polar
``antiferrodistortive'' instability, and that the relative strength of
these two instabilities depends strongly on the unit cell volume
\cite{Bhattacharjee/Bousquet/Ghosez:2009,Rondinelli/Eidelson/Spaldin:arXiv}. For
larger volumes ({\it i.e.} BaMnO$_3$) the ferroelectric instability
becomes dominant. Thus, even though BaMnO$_3$ is not stable in the
cubic (or in the orthorhombically distorted) perovskite structure (it
crystallizes in a hexagonal structure), this opens up the possibility
to stabilize the corresponding ferroelectric phase by using epitaxial
constraints, {\it i.e.} using thin film growth techniques.

Apart from these investigations into possible mechanisms for
ferroelectricity that are compatible with the simultaneous presence of
magnetic order, first principles calculations have also been used to
rationalize experimental observations, investigate possible mechanisms
for coupling between the electric polarization and the magnetic order,
and to design new multiferroic and magneto-electric materials. In the
following we will highlight some of these calculations, in particular
the work related to one of the most prominent multiferroic materials:
bismuth ferrite.

\subsection{First principles calculations for BiFeO$_3$ and related work}

BiFeO$_3$ (BFO) is one of the most studied (probably \emph{the} most
studied) multiferroic material. BFO is known to be multiferroic (or
more precisely: AFM and ferroelectric) already since the early 1960s
\cite{Kiselev/Ozerov/Zhdanov:1963}. However, for a long time it was
not considered as a very promising material for applications, since
the electric polarization was believed to be rather small
\cite{Teague/Gerson/James:1970} and the AFM order does not lead to a
net magnetization
\cite{Fischer_et_al:1980,Sosnowska/Peterlin-Neumaier/Streichele:1982}.

This has changed drastically, following a publication in Science in
2003 (Ref.~\cite{Wang_et_al:2003}), which to great extent has
triggered the intensive experimental and theoretical/computational
research on BFO during the last 5--6 years. In this study, a large
spontaneous electric polarization in combination with a substantial
magnetization was observed above room temperature in thin films of BFO
grown epitaxially on SrTiO$_3$ substrates. The presence of both
magnetism and ferroelectricity above room temperature, together with
potential coupling between the two order parameters, makes BFO the
prime candidate for device applications based on multiferroic
materials.

Whereas the large electric polarization was later confirmed
independently, and explained by first principles calculations, the
origin of the strong magnetization reported in \cite{Wang_et_al:2003}
is still unclear and, to the best of our knowledge, it has never been
reproduced in an independent study. It is generally assumed that the
magnetization reported in Ref.~\cite{Wang_et_al:2003} is related to
extrinsic effects such as defects or small amounts of impurity phases.

The large electric polarization, which appeared to be at odds with
bulk single crystal measurements from 1970
\cite{Teague/Gerson/James:1970}, was originally assumed to be due to
epitaxial strain, which results from the lattice constant mismatch
between BFO and the substrate material SrTiO$_3$. It is known that
epitaxial strain can have drastic effects on the properties of thin
film ferroelectrics. For example, it can lead to a substantial
enhancement of electric polarization and can even induce
ferroelectricity at room temperature in otherwise non-ferroelectric
SrTiO$_3$ \cite{Choi_et_al:2004,Haeni_et_al:2004}.

In the following we illustrate how first principles calculations have
been instrumental in clarifying the origin of both polarization and
magnetization in thin film BFO, by showing that the large electric
polarization found in the thin films is in fact intrinsic to
unstrained bulk BFO and that, in contrast to many other
ferroelectrics, epitaxial strain has only a minor effect in this
material.

\subsubsection{Electric polarization of bulk BFO and the effect of
  epitaxial strain}

According to the so-called ``Modern theory of polarization'', the
electric polarization of a bulk periodic system is defined via the
Berry phase of the corresponding wavefunctions
\cite{King-Smith/Vanderbilt:1993,Resta:1994}. Since this geometrical
phase is only well defined modulo $2\pi$, the polarization is only
well-defined modulo so-called ``polarization quanta'', given by
$\Delta\vec{P}_{0}^{(i)} = \frac{fe}{\Omega} \vec{a}_i$, where $e$ is
the electronic charge, $\vec{a}_i$ a primitive lattice vector
($i=1,2,3$), $\Omega$ the unit cell volume, and $f$ is a spin
degeneracy factor ($f=2$ for a non-spinpolarized system, $f=1$ for a
spin-polarized system). If the expression for the polarization is
recast as a sum over ``Wannier centers''
\cite{King-Smith/Vanderbilt:1993}, a translation of one of the
occupied Wannier states from one unit cell to the next corresponds to
a change in polarization by exactly one ``quantum''. The
multivaluedness thus reflects the arbitrary choice of basis vectors
when describing an infinite periodic structure.

In spite of this multivaluedness of the bare polarization for a
specific atomic configuration, differences in polarization are well
defined quantities, provided the corresponding configurations can be
transformed into each other in a continuous way and the system remains
insulating along the entire ``transformation path'' \cite{Resta:1994}.

In particular, the spontaneous polarization of a ferroelectric
material is defined as half the difference in polarization between two
oppositely polarized states, or equivalently, as the difference in
polarization between the ferroelectric structure and a suitable
centrosymmetric reference configuration. In order to calculate the
spontaneous polarization one therefore has to perform a series of
calculations for different configuration between the ferroelectric
state and the centrosymmetric reference structure. If the change in
polarization between two such configurations is much smaller than the
polarization quantum, then the corresponding difference can be clearly
identified and the full change in polarization along the
transformation path, {\it i.e.} the spontaneous polarization, can be
determined.

The application of this procedure to calculate the spontaneous
polarization of BFO is complicated by the following two features: {\it
  i)} the polarization quantum for a spin-polarized system is only
half that for a similar nonmagnetic system, and {\it ii)} due to the
underestimation of the local spin splitting for Mott-Hubbard
insulators within the standard local spin-density approximation
(LSDA), BFO becomes metallic for the less distorted reference
configurations within LSDA.

These problems have been overcome in Ref.~\cite{Neaton_et_al:2005} by
using the LSDA+$U$ method
\cite{Anisimov/Zaanen/Andersen:1991,Anisimov/Aryatesiawan/Liechtenstein:1997}
to calculate the electronic structure of BFO in various configurations
along the transformation path from the fully distorted $R3c$ structure
to the centrosymmetric cubic perovskite ($Pm\bar{3}m$)
structure. Within the LSDA+$U$ method the local $d$-$d$ exchange
splitting is enhanced by the Hubbard $U$ and BFO stays insulating even
in the undistorted cubic perovskite structure (for $U$ values
$U_\text{eff} = U - J =$ 2--4~eV \cite{Neaton_et_al:2005}).

\begin{figure}
\centerline{\includegraphics[width=0.85\textwidth]{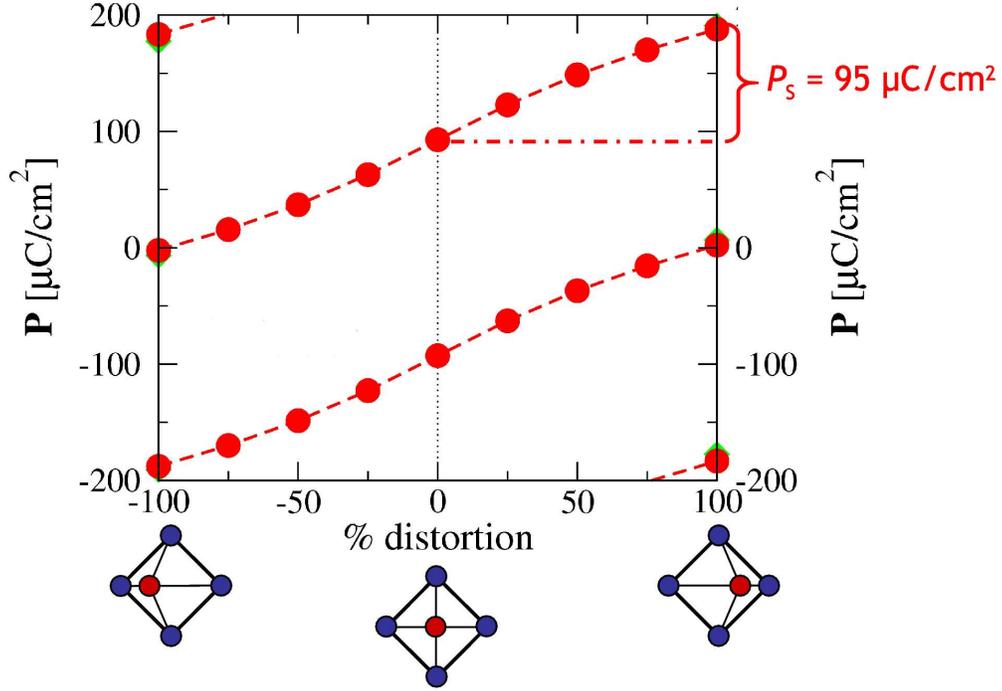}}
\caption{\label{fig:bfo-pol} Evolution of the polarization $P$ along
  the transformation path from a negatively polarized state ($-100$~\%
  distortion), through a centrosymmetric reference configuration (0~\%
  distortion), to a positively polarized state ($+100$~\%
  distortion). Red circles correspond to the LSDA+$U$ calculation with
  $U_\text{eff}=2$~eV, green diamonds indicate the LSDA result for the
  fully polarized states. Different values of $P$ for fixed amount of
  distortion are separated by the polarization quantum $\Delta
  P^{(111)}_0 = 186$~$\mu$C/cm$^2$. The spontaneous polarization $P_s$
  is given by the difference in polarization between the fully
  distorted and the undistorted configuration for an arbitrary branch
  of the bare polarization. Note: the systematic sketches at the
  bottom do not correspond to the actual crystal structure of BFO.}
\end{figure}

Figure \ref{fig:bfo-pol} shows the evolution of the electric
polarization with varying degree of distortion between two oppositely
polarized states calculated for $U_\text{eff} = 2$~eV. The LSDA
results are included for $\pm 100$~\% distortion. The fact that the
corresponding symbols (green diamonds) can barely be recognized behind
the red circles that indicate the LSDA+$U$ results shows that the
value of the bare polarization is rather insensitive to the exact
value of $U_\text{eff}$. It can be seen that different values of $P$
corresponding to the same amount of distortion are separated by the
polarization quantum along (111), $\Delta P_0^{(111)} =
\frac{e}{\Omega} (\vec{a}_1 + \vec{a}_2 + \vec{a}_3)$, where
$\vec{a}_{1,2,3}$ are the primitive lattice vectors of the
rhombohedral $R3c$ structure. As indicated, the spontaneous
polarization $P_s$ can be obtained as the difference between the fully
distorted and the undistorted configuration for an arbitrary
``branch'' of the bare polarization.

From these calculation a spontaneous polarization of bulk BFO of $\sim
95$~$\mu$C/cm$^2$ has been obtained. This is an order of magnitude
larger than what was previously believed to be the case, based on the
measurements in Ref.~\cite{Teague/Gerson/James:1970}, and even exceeds
the polarization of typical prototype ferroelectrics such as
BaTiO$_3$, PbTiO$_3$, or PbZr$_{0.5}$Ti$_{0.5}$O$_3$ (PZT). Variation
of $U_\text{eff}$ within reasonable limits changes the calculated
value for the electric polarization by only $\sim \pm
5$~$\mu$C/cm$^2$, {\it i.e.} the large value of the polarization is
rather independent from the precise value of the Hubbard
parameter. This is consistent with the assumption that the transition
metal $d$ states do not play an active role for the ferroelectric
instability in BFO. The calculated large spontaneous polarization for
bulk BFO is also consistent with the large ionic displacements in the
experimentally observed $R3c$ structure of BFO (see
Fig.~\ref{fig:structures}c), compared to an appropriate
centrosymmetric reference configuration. Recently, the large
polarization of $\sim 100$~$\mu$C/cm$^2$ along (111) for bulk BFO has
also been confirmed experimentally by new measurements on high-quality
single crystals \cite{Lebeugle_et_al:2007}.

\begin{figure}
\centerline{\includegraphics[width=0.6\textwidth]{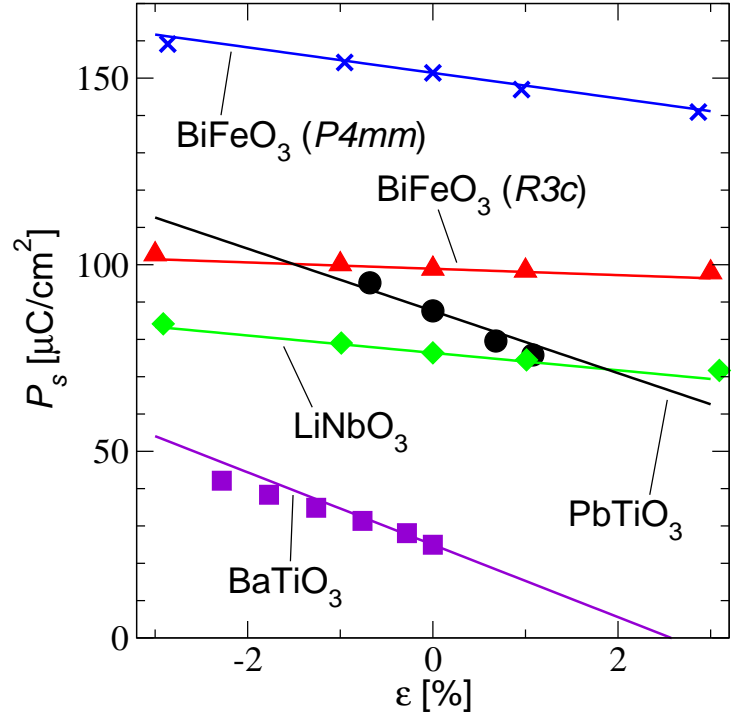}}
\caption{Dependence of the spontaneous polarization $P_s$ on epitaxial
  strain $\epsilon$ for BFO in two different structural modifications
  and some other (non-magnetic) ferroelectrics. Symbols correspond to
  results from first principles calculations for strained unit cells
  (data for BaTiO$_3$/PbTiO$_3$ is taken from
  \cite{Neaton/Hsueh/Rabe:2002}/\cite{Bungaro/Rabe:2004}), lines are
  obtained from the calculated bulk linear response functions (see
  \cite{Ederer/Spaldin_PRL:2005}). Note that the epitaxial constraint
  for all systems is applied in the plane perpendicular to the
  polarization, {\it i.e.} (001) for BaTiO$_3$, PbTiO$_3$, and
  $P4mm$-BiFeO$_3$, and (111) for LiNbO$_3$ and $R3c$-BiFeO$_3$.}
\label{fig:strain}
\end{figure}

Effects of epitaxial strain can be assessed from first principles by
performing bulk calculations for a strained unit cell, where the
lattice constant within a certain lattice plane (corresponding to the
orientation of the substrate surface) is constrained, whereas the
lattice constant in the perpendicular direction as well as all
internal structural parameters are allowed to relax. Such calculations
have been performed for BFO corresponding to a (111) orientation of
the substrate \cite{Ederer/Spaldin_2:2005}. In this case the $R3c$
symmetry of the bulk structure is conserved and the epitaxial
constraint is applied in the lattice plane perpendicular to the
polarization direction. It was found that the sensitivity of the
electric polarization to strain is surprisingly weak in BFO, much
weaker than in other well-known ferroelectrics
\cite{Ederer/Spaldin_2:2005} (see Fig.~\ref{fig:strain}). A systematic
comparison of the strain dependence in various ferroelectrics,
including BFO in both the $R3c$ and a hypothetical tetragonal phase
with $P4mm$ symmetry, has been performed in
Ref.~\cite{Ederer/Spaldin_PRL:2005} (see Fig.~\ref{fig:strain}). It
was shown that the effect of epitaxial strain for all investigated
systems can be understood in terms of the usual bulk linear response
functions and that both strong and weak strain dependence can occur.

Systematic calculations corresponding to a (001) orientation of the
substrate, the one that is most often used experimentally, have not
been performed so far. Since the epitaxial constraint in this case
breaks the rhombohedral symmetry of the bulk structure, the
corresponding strained unit cell has a lower symmetry with more free
parameters than in the (111)-strained case. Nevertheless, the effect
of such a monoclinic strain on the ferroelectric polarization in BFO
has been investigated by performing calculations for a set of lattice
parameters derived from representative experimental data. Due to the
lower symmetry, the polarization in this case is slightly rotated away
from the (111) direction, but the overall magnitude remains nearly
unchanged compared to the unstrained case. From this it was concluded
that the polarization in BFO is generally rather insensitive to
epitaxial strain, and that the large polarization measured in thin
films is basically the same as in the corresponding bulk
system. Indeed, the polarization of $\sim 60$~$\mu$C/cm$^2$ reported
in Ref.~\cite{Wang_et_al:2003} for a (001) oriented thin film agrees
well with the corresponding projection of the calculated bulk value
(which is oriented along the (111) direction), and polarization
measurements for BFO films with different substrate orientations
((001), (101), and (111)) can all be understood by assuming that the
polarization vector in all cases points essentially along (111) and
has approximately the same length \cite{Li_et_al:2004}. More recently,
systematic experimental investigations of the strain effect in
epitaxial BFO films have been undertaken by comparing results of BFO
films with different thicknesses, which have confirmed the predicted
weak strain dependence of the polarization in BFO
\cite{Kim_et_al:2008}

Finally, it should be noted that Ref.~\cite{Wang_et_al:2003} also
contains results of first principles calculations for the electric
polarization of two structural variants of BFO: the rhombohedral bulk
structure with $R3c$ space group, and a hypothetical tetragonal
structure with $P4mm$ symmetry, based on the lattice parameters found
in the thin film samples. At that time it was assumed that such a
tetragonal phase is stabilized in epitaxial thin films and that the
difference in polarization observed in thin films compared to bulk BFO
was due to a large difference in polarization between the two
different structural modifications. However, the DFT results presented
in Ref.~\cite{Wang_et_al:2003} were not conclusive, since only the
bare polarization for the two different structures was reported and
not the spontaneous polarization that is measured in the corresponding
``current-voltage'' switching experiments.

In fact, it is indeed possible that a different phase is stabilized in
thin films, which can then lead to more significant changes of
ferroelectric and magnetic properties compared to bulk BFO. However,
it is important to distinguish between the simple case of a somewhat
distorted version of the rhombohedral bulk structure and a truly
different phase, which would for example be characterized by a
different oxygen octahedra tilt pattern or a different number of
formula units contained in the crystallographic unit cell.

Calculations presented in Ref.~\cite{Ederer/Spaldin_PRL:2005} (see
also \cite{Ricinschi/Yun/Okuyama:2006}) show that if BFO is
constrained to tetragonal $P4mm$ symmetry (with no octahedral tilts
and only one formula unit per unit cell) it develops a
''super-tetragonality'' with $c/a$ ratio of 1.27 and a giant electric
polarization of $P_s \approx 150$~$\mu$C/cm$^2$. A polarization of
this magnitude has indeed been found in some highly strained films
with $c/a$ ratios between 1.2--1.3
\cite{Yun_et_al:2004,Ricinschi/Yun/Okuyama:2006}, whereas many other
experimental reports of ``tetragonal'' BFO films with smaller $c/a$
ratio also exist. These reports should be regarded with some caution,
since the structural characterization of thin films is usually
restricted to the measurement of lattice constants and of angles
between certain crystallographic directions. A full characterization
of ionic distortions (including octahedral tilt patterns etc.) is
generally not possible for thin films, and first principles
calculations can therefore play an important role in clarifying open
questions about the exact thin film structure of BFO. In principle, if
one tries to epitaxially match the rhombohedral bulk structure of BFO
on a square lattice substrate plane, one can expect to obtain a
monoclinically distorted version of the BFO bulk structure. However,
since the rhombohedral angle in bulk BFO is very close to 60$^\circ$,
the value that corresponds to an underlying cubic lattice, the
monoclinic distortion can be rather small, and the thin films might
appear essentially tetragonal.

\subsubsection{Weak ferromagnetism in thin film BFO and coupling between
  the various order parameters}

In addition to these structural studies, DFT calculations have also
been used to investigate the magnetic properties of BFO, in particular
the possible origin for the significant magnetization reported in
Ref.~\cite{Wang_et_al:2003}. Bulk BFO is known to exhibit ``G-type''
AFM ordering \cite{Fischer_et_al:1980}, {\it i.e.} the magnetic moment
of each Fe cation is antiparallel to that of its nearest
neighbors. Superimposed to this G-type magnetic order a long-period
cycloidal modulation is observed, where the AFM order parameter
$\vec{L} = \vec{M}_1 - \vec{M}_2$, defined as the difference between
the two sublattice magnetizations $\vec{M}_{1,2}$, rotates within the
(110) plane with a wavelength of $\sim
620$~\AA\ \cite{Sosnowska/Peterlin-Neumaier/Streichele:1982}.

\begin{figure}
\centerline{\includegraphics[width=0.8\textwidth]{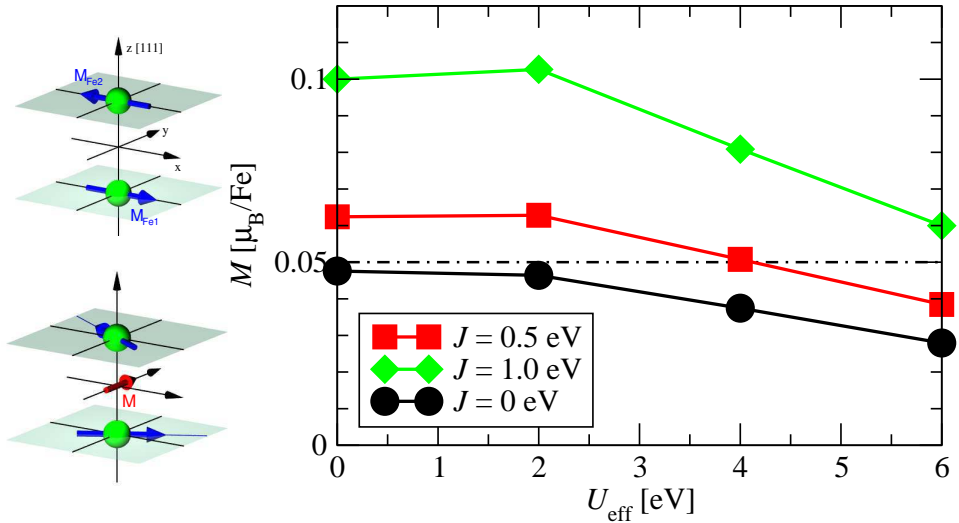}}
\caption{Dependence of the weak magnetization in BFO on the LSDA+$U$
  parameters $U_\text{eff} = U - J$ and $J$. The dash-dotted line
  represents the reported value of 0.05~$\mu_\text{B}$/Fe. The
  sketches on the left side illustrate how the canting of the two AFM
  sublattice magnetizations, represented by the magnetic moments
  $M_\text{Fe1}$ and $M_\text{Fe2}$ of the two Fe cations in the
  primitive unit cell, gives rise to the net magnetization $M$.}
\label{BFO-wfm}
\end{figure}

Calculations for bulk BFO show a very strong and dominant AFM nearest
neighbor interaction \cite{Baettig/Ederer/Spaldin:2005}, in agreement
with the observed G-type magnetic order and the rather high N{\'e}el
temperature of $\sim 600$~K. In addition, the magnetocrystalline
anisotropy has been calculated, and a preferred orientation of the Fe
magnetic moments perpendicular to the polar [111] direction has been
found \cite{Ederer/Spaldin:2005}. Within the (111) plane a 12-fold
degeneracy remains, leading to an effective ``easy-plane'' geometry
for the magnetic moments. For an orientation of the AFM order
parameter $\vec{L}$ within this (111) plane, \emph{weak
  ferromagnetism} is symmetry-allowed, {\it i.e.} a small canting of
the two AFM sublattice magnetizations can occur, which results in a
net magnetization \cite{Dzyaloshinskii:1957}. Indeed, if spin-orbit
coupling is included in the calculation (while the cycloidal
modulation is neglected), a small canting of the magnetic moments is
obtained \cite{Ederer/Spaldin:2005}. The magnitude of the resulting
magnetization depends on the choice of the Hubbard $U$ and the Hund's
rule parameter $J$, but for reasonable values of $U_\text{eff} = U-J$
the magnetization is around 0.05~$\mu_\text{B}$/Fe cation (see
Fig.~\ref{BFO-wfm}). This value of the magnetization agrees quite well
with various thin film measurements
\cite{Eerenstein_et_al:2005,Bai_et_al:2005,Bea_et_al:2007}, but is
significantly smaller than what was originally reported in
Ref.~\cite{Wang_et_al:2003}. It has to be pointed out that no
magnetization is observed in bulk BFO, where the presence of the
cycloidal modulation effectively cancels any net magnetic moment. If
the cycloidal modulation is suppressed, either by applying a strong
magnetic field \cite{Popov_et_al:1993} or by chemical substitution
\cite{Sosnowska_et_al:2002} a small magnetization appears, with
comparable magnitude to the computational result. It is generally
assumed that the cycloidal rotation of the AFM order parameter is also
suppressed in thin films, likely due to enhanced anisotropy, and that
the small magnetization observed in the thin films is due to weak
ferromagnetism. This is supported by a neutron diffraction study on
BFO films, which could not find the satellite peaks associated with
the cycloidal modulation \cite{Bea_et_al:2007}.

Furthermore, first principles studies addressing the effect of
epitaxial strain and the presence of oxygen vacancies did not find a
significant increase in magnetization \cite{Ederer/Spaldin_2:2005},
and it is therefore likely that the large magnetization reported in
\cite{Wang_et_al:2003} is due to other defects or small amounts of
impurity phases.

The appearance of weak ferromagnetism in thin films of BFO leads to
the question of whether this small magnetization is coupled to the
electric polarization, {\it i.e.} whether it can be manipulated by
applying external electric fields. Indeed, the absence of an inversion
center located at the midpoint between two interacting magnetic
moments is crucial to produce a non-vanishing
\emph{Dzyaloshinskii-Moriya (DM) interaction}, which has been
identified as the microscopic mechanism responsible for the magnetic
moment canting in weak ferromagnets \cite{Moriya:1960}. Thus,
inversion symmetry breaking can cause both weak ferromagnetism and
ferroelectricity, suggesting possible cross-correlations between these
two properties. First principles calculations have been used to
explore this possibility for magnetization-polarization coupling in
BFO \cite{Ederer/Spaldin:2005} and in BaNiF$_4$
\cite{Ederer/Spaldin:2006}. It was found that in BFO the DM
interaction is caused by a non-polar antiferrodistortive mode, not by
the polar distortion, and therefore the weak ferromagnetism in BFO is
not controlled by the spontaneous polarization and cannot be switched
using an electric field \cite{Ederer/Spaldin:2005}. In contrast, in
BaNiF$_4$, it is indeed the polar distortion that creates a DM
interaction, but the symmetry is such that no net magnetization
results. Instead, a secondary (weak) AFM order parameter is induced in
addition to the distinctly different primary AFM order
\cite{Ederer/Spaldin:2006}. Only recently, a material has been
suggested, based on a combination of first principles calculations and
symmetry considerations, that fulfills all requirements for
``ferroelectrically-induced weak ferromagnetism''
\cite{Fennie:2008}. The corresponding material, $R3c$ structured
FeTiO$_3$, is closely related to BFO in that it has the same overall
structural symmetry, but with the magnetic Fe cations located on the
perovskite $A$ site instead of the perovskite $B$ site as in BFO. It
is this difference in the local site symmetry of the magnetic cation,
that is crucial for the coupling between the spontaneous polarization
and the weak magnetization
\cite{Fennie:2008,Ederer/Fennie:2008}. Experimental work is currently
underway to validate this theoretical prediction.

\subsubsection{Designing new multiferroics and new functionalities}

The prediction of FeTiO$_3$ as a possible candidate for electric field
switchable weak ferromagnetism, is one example for attempts to design
new materials with novel or more favorable magneto-electric properties
based on first principles electronic structure calculations.

Another example is the design of a material that allows for
magneto-electric phase control \cite{Fennie/Rabe_2:2006}. Calculations
for the rare-earth magnet EuTiO$_3$ showed that this material exhibits
a soft infrared-active, {\it i.e.} polar, phonon mode that becomes
unstable if the material is epitaxially strained. In addition, due to
strong spin-phonon coupling in this material, the instability is more
pronounced for ferromagnetic ordering of the Eu spins than for the
case of an AFM arrangement. Since the ground state magnetic structure
for the lower strain region is AFM, it was suggested that a phase
transition from a non-polar AFM phase into a
ferroelectric-ferromagnetic phase can be induced by applying a strong
magnetic field, if the material can be prepared in thin films with a
compressive epitaxial strain of around 1~\% \cite{Fennie/Rabe_2:2006}.

In addition, attempts have been made to design materials that combine
strong ferroelectric polarization with a large magnetization above
room temperature. If such a material would also exhibit pronounced
coupling effects between polarization and magnetization, which ideally
would allow to switch the polarization via a magnetic field or vice
versa, then this would probably create a similar excitement as finding
a room temperature superconductor. Unfortunately, at the moment no
multiferroic that exhibits all these properties is known (similarly,
no room temperature superconductor is known at present).

A suggestion for a material combining large polarization and large
magnetization has been made in Ref.~\cite{Baettig/Spaldin:2005}. First
principles calculations predict, that if half of the Fe$^{3+}$ cations
in BFO are replaced by Cr$^{3+}$ cations in a checkerboard-like
ordered arrangement, then the resulting material Bi$_2$FeCrO$_6$ is
stable in a rhombohedral structure similar to BFO with a spontaneous
ferroelectric polarization of around 80~$\mu$C/cm$^2$ and a
magnetization of 2~$\mu_\text{B}$ per formula unit. The magnetization
in this case results from a ferri-magnetic arrangement, where the
magnetic moments of the Cr cations are antiparallel to those of the Fe
cations. A subsequent study of the strength of the magnetic coupling
in the series of compounds BiFeO$_3$-Bi$_2$FeCrO$_6$-BiCrO$_3$ has
found that the N{\'e}el-temperature in Bi$_2$FeCrO$_6$ is unlikely to
be above room temperature \cite{Baettig/Ederer/Spaldin:2005}, but
nevertheless several attempts have been made to synthesize the
corresponding material
\cite{Nechache_et_al:2006,Suchomel_et_al:2007,Kim_et_al:2007}. The
synthetic challenge here, is to achieve the required checkerboard-type
ordering of Fe and Cr cations on the $B$ sites of the underlying
perovskite structure, which might be possible by utilizing
layer-by-layer growth on a (111)-oriented substrate.

\subsection{Perspectives for future studies of proper multiferroics}

The examples discussed so far show that first principles calculations
have proven not only to be useful for rationalizing experimental
observations and identifying different mechanisms for ferroelectricity
that can be found in multiferroic materials, but also to facilitate
quantitative predictions of new materials and novel effects in proper
magnetic ferroelectrics. Future applications of {\it ab initio}
methods in the design of new materials and in calculating the expected
properties of these materials are therefore expected to continue to
have a significant impact on the overall progress of this field.

In particular, a material with large magnetization and large
polarization above room temperature is still elusive. From the current
point of view there is no fundamental reason why such a material
should not exist, and creative ideas on how to circumvent the
limitations and restrictions of materials chemistry that have been
encountered so far are still highly desirable.

Another area where DFT will undoubtedly have (and already has) a
substantial impact, is the study of artificial heterostructures
consisting of a combination of magnetic and ferroelectric materials
\cite{Ramesh/Spaldin:2007}. Examples of computational work in that
direction that have already appeared include the study of artificial
tri-layered superlattices of different magnetic and nonmagnetic oxides
\cite{Hatt/Spaldin:2007} and the investigation of polarization effects
at the interface between a ferromagnetic metal and a ferroelectric
insulator \cite{Duan/Jaswal/Tsymbal:2006}.

\section{Improper Multiferroics}

In the beginning of this section, we will focus on the origin of
ferroelectricity in the so-called ``Improper multiferroics" (IMF),
outlining a few differences with respect to the more conventional
``proper" multiferroics discussed so far.

As pointed out in the previous sections, in displacive ferroelectric
materials (such as prototypical perovskite-like BaTiO$_3$ or
multiferroic BiFeO$_3$), due to strong covalency effects, the relative
displacement of the anionic sublattice with respect to the cationic
sublattice gives rise to a spontaneous and switchable polarization,
which is the (primary) order parameter in the ferroelectric
transition.  On the other hand, in IMF, the primary order parameter of
the phase transition is related to electronic ({\em i.e.} spin,
charge, or orbital) degrees of freedom
\cite{Cheong/Mostovoy:2007}. The important thing is that the resulting
electronic order lacks inversion symmetry (IS), therefore opening the
way to ferroelectricity. Therefore, polarization occurs as a
by-product of the electronic phase transition and can be described as
a ``secondary" order parameter. As a consequence, {\em i}) even the
state with ions pinned in centrosymmetric positions can show a finite
(purely electronic) polarization; {\em ii}) the ions can ``react" to
the non-centrosymmetric charge-redistribution by displacing, so as to
give a (more traditional) ionic contribution to the total
polarization. In order to push ahead with the comparison between
proper and improper multiferroics, one can say that ferroelectricity
in IMF is driven by ``correlation" effects (as related to spin or
charge arrangements), at variance with the previously mentioned case
of standard ferroelectrics where it is mostly driven by covalency. In
IMF where polarization is magnetically-induced, it is reasonable to
expect a strong coupling between magnetic and ferroelectric
properties, since the two dipolar and magnetic orderings share the
same origin and occur at the same temperature.

\begin{figure}
\resizebox{160mm}{!}
{\includegraphics[angle=90]{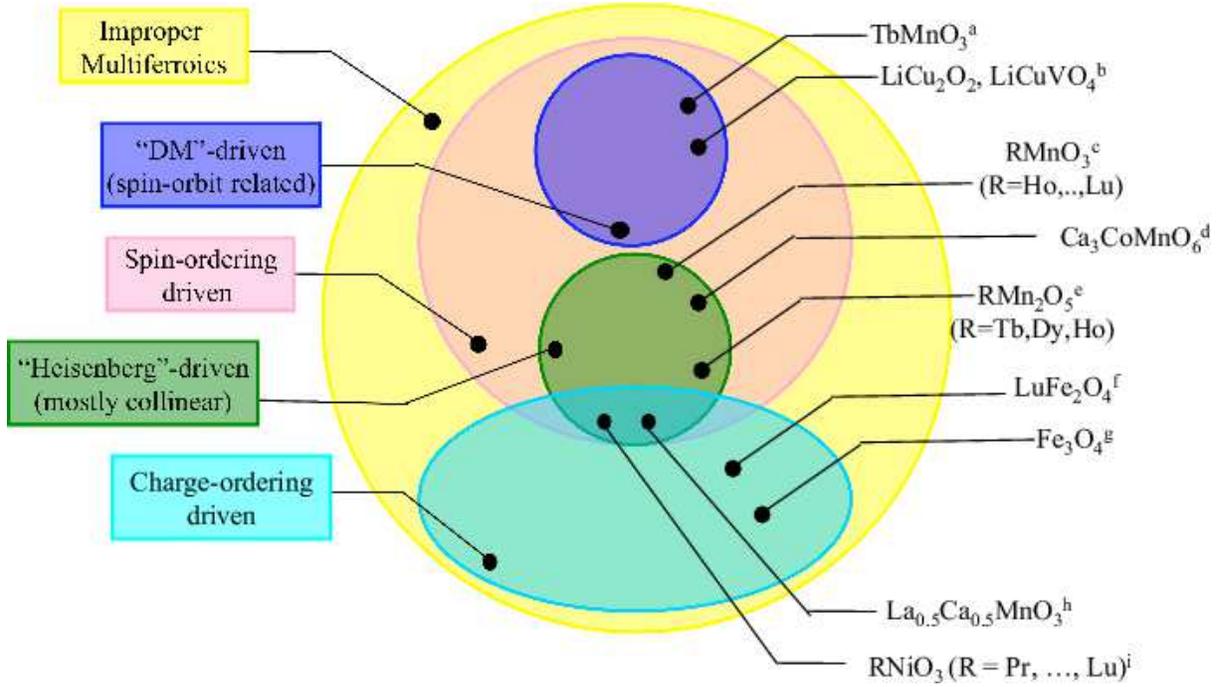}}
\caption{\label{fig:imf} Schematic classification of IMF, in terms of
  different mechanisms (left side) and compounds (right side). The
  (non comprehensive) list includes a few materials which were studied
  by first-principles (see related references: a)
  Ref.~\cite{xiang,malashevic}, b) Ref.~\cite{whangbolicuo}, c)
  Ref.~\cite{prlslv,prbkyama}, d) Ref.~\cite{ccmo1,ccmo2}, e)
  Ref.~\cite{lixinhe,prlgianluca}, f) Ref.~\cite{lufeo}, g)
  Ref.~\cite{alexe}, h) Ref.~\cite{gglcmo}, i) Ref.~\cite{ggni}).}
\end{figure}

In Fig.~\ref{fig:imf} we schematically classify IMF on the basis of
the different mechanisms to induce ferroelectricity that have been
proposed so far.  We would like to point out that what we present in
the following is a non-exhaustive list of the IMF materials and
related mechanisms. In fact, IMF represent a quickly evolving field:
new materials and/or novel mechanisms are proposed on a monthly or
even weekly basis. With no doubt, we therefore expect in the near
future this classification to become richer in compounds and to expand
as far as mechanisms are concerned.

In Fig.~\ref{fig:imf} IMF are divided in two main classes: those where
ferroelectricity is driven by spin-order ({\em i.e.} where the
``magnetic" arrangement breaks IS) and those where it is driven by
charge-order ({\em i.e.} where the charge-disproportionation leads to
a non-centrosymmetric arrangement). In turn, the magnetically-induced
ferroelectricity can occur in two different ways: {\em i}) the first
and most studied case where a non-collinear spin-spiral occurs and the
IS-breaking arises due to a spin-orbit related mechanism in the
DM-like antisymmetric exchange term \cite{mostovoy,nagaosa,ivansoc};
{\em ii}) the case of (mostly collinear) AFM spins where the
IS-breaking occurs in the Heisenberg-like symmetric exchange-term
\cite{sergienko,prlslv}.

Along with the classification of IMF, we show in Fig.~\ref{fig:imf} a
few links to IMF materials for which {\em ab-initio} studies have been
reported in the literature.

Chronologically, the recent interests towards IMF were boosted by the
discovery of ferroelectricity in TbMnO$_3$ and of the control of the
polarization direction achieved via an applied magnetic field
\cite{kimura}. However, the {\em ab-initio} simulations for TbMnO$_3$
came much later \cite{malashevic,xiang}, due to the complexity in the
related simulations: advanced capabilities (such as non-collinear
magnetism and spin-orbit coupling) are needed to reproduce the
observed tiny effects, which implicitly requires a high precision in
terms of numerical parameters in the calculations. In the {\em
  ab-initio} field, the first IMF to be studied were collinear
antiferromagnets, such as TbMn$_2$O$_5$ \cite{lixinhe} and HoMnO$_3$
\cite{prlslv}. Since the latter will be described in detail in
Sec.~\ref{E-type}, we will now briefly discuss the first one.  The
class of manganites often labeled as ``1-2-5'' from the stoichiometry
of rare-earth, transition metal, and oxygen, respectively, is an
actively studied set of IMF. Despite some non-collinearity and
non-commensurability effects, most of the mechanisms behind
multiferroicity can be described through simulations with
non-centrosymmetric collinear spin arrangement using a relatively
small supercell. The suggested polarization was of the order of 1 $\mu
C/cm^2$ and the polarization was reversed by changing the
spin-orientation in the unit cell, providing evidence for the magnetic
origin of ferroelectricity in TbMn$_2$O$_5$. Within the same class of
materials, HoMn$_2$O$_5$ was studied in Ref.~\cite{prlgianluca}: the
main and new result of that work was that the ionic and electronic
contributions were strongly dependent on the value of the Hubbard $U$
parameter used in a LSDA+$U$ approach, pointing to the important role
of correlation effects in 1-2-5 manganites.

Within the spin-spiral class of IMF, Li-Copper-based oxides were the
first compounds to be studied from first-principles
\cite{whangbolicuo}: upon switching-on spin-orbit coupling, the
calculated polarization was rather small (of the order of tens or
hundreds of $\mu$C/cm$^2$, depending on whether ionic relaxations were
included or not in the simulations). Shortly later, the prototypical
case of TbMnO$_3$ was published in two important papers (one following
the other in Phys. Rev. Lett.), Refs.~\cite{malashevic,xiang}. It was
shown that the purely electronic contribution ({\em i.e.} evaluated by
switching on spin-orbit but keeping the ions frozen into their
paramagnetic centrosymmetric configuration) was much smaller than the
ionic contribution ({\em i.e.} evaluated by relaxing the ions).  In
the TbMnO$_3$ case, the order of magnitude of the {\em ab-initio}
polarization was found to be in excellent agreement with experiments
\cite{kimura}.  Remarkably, at the time of publication, the sign of
polarization obtained within DFT was opposite with respect to
experiments; indeed, it later turned out \cite{sign} that the
discrepancy was due to a misunderstanding in the conventions of the
experimental settings and an excellent agreement between theory and
experiments could be finally obtained.

Within the field of charge-order-induced ferroelectricity, a prototype
has emerged: the triangular mixed-valence iron-oxide, LuFe$_2$O$_4$
\cite{ikeda}. There, the frustrated charge-ordering is such as to lack
centrosymmetry: in each FeO bilayer, there is an alternation of iron
atoms, with Fe$^{2+}$:Fe$^{3+}$ ratios of 2:1 and 1:2, therefore
giving rise to a polarization within each bilayer. The polarization
estimated from first-principles is very large (of the order of 10 $\mu
C/cm^2$ in the bilayer). However, some controversy exists for that
material, since it is questioned whether the stacking of the bilayers
is such as to produce net ferroelectricity \cite{lufeo} or a global
antiferroelectricity with no net polarization \cite{lufeonew}. More
work (both from theory and from experiments) will be needed in that
respect.

Recently, another collinear compound has been studied, Ca$_3$CoMnO$_6$
\cite{ccmo1,ccmo2}. The main {\em ab-initio} findings were: {\em i}) a
large Co orbital moment, which renders the system similar to an
Ising-like chain, with alternating trigonal prismatic Co$^{2+}$ and
octahedral Mn$^{4+}$ sites in the spin chain; {\em ii}) a large
calculated polarization (about 1.7 $\mu$C/cm$^2$), caused by a
significant exchange-striction combined with a peculiar $\uparrow
\uparrow \downarrow \downarrow$ spin configuration.

Given this general background, in the following sections we will
present some examples of {\em ab-initio} calculations for IMF. In
closer detail, we will discuss rare-earth manganites ({\it cfr.}
Sec.~\ref{E-type}) \cite{prlslv,prbkyama} and hole-doped manganites
({\it cfr.} Sec.~\ref{hdm}) \cite{gglcmo} as examples of AFM materials
where the spin-arrangements break inversion symmetry, with
polarization being due to Heisenberg-like mechanisms. We will conclude
the section by discussing some perspectives and open issues in the
field.

In what follows, we will mainly show the results of DFT simulations
performed using the Vienna Ab-initio Simulation Package (VASP)
\cite{vasp} and the generalized gradient approximation \cite{pbe} to
the exchange-correlation potential. For the construction of the
Wannier functions, we used the Full-potential Linearized Augmented
Plane-Wave (FLAPW) \cite{flapw} code in the FLEUR implementation
\cite{fleur}. For a better treatment of correlation effects, the
so-called LSDA+$U$ approach
\cite{Anisimov/Aryatesiawan/Liechtenstein:1997} (with $U = 4$~eV and
$J = 0.9$~eV) was used in the case of hole-doped manganites. For
further technical details, as far as computational or structural
parameters are concerned, we refer to our original publications
\cite{prlslv,prbkyama,gglcmo}.

\subsection{Highlights on Improper Multiferroics}

\subsubsection{E-type rare-earth ortho-manganites}
\label{E-type}

\begin{figure}
\resizebox{160mm}{!}  
{\includegraphics[angle=90]{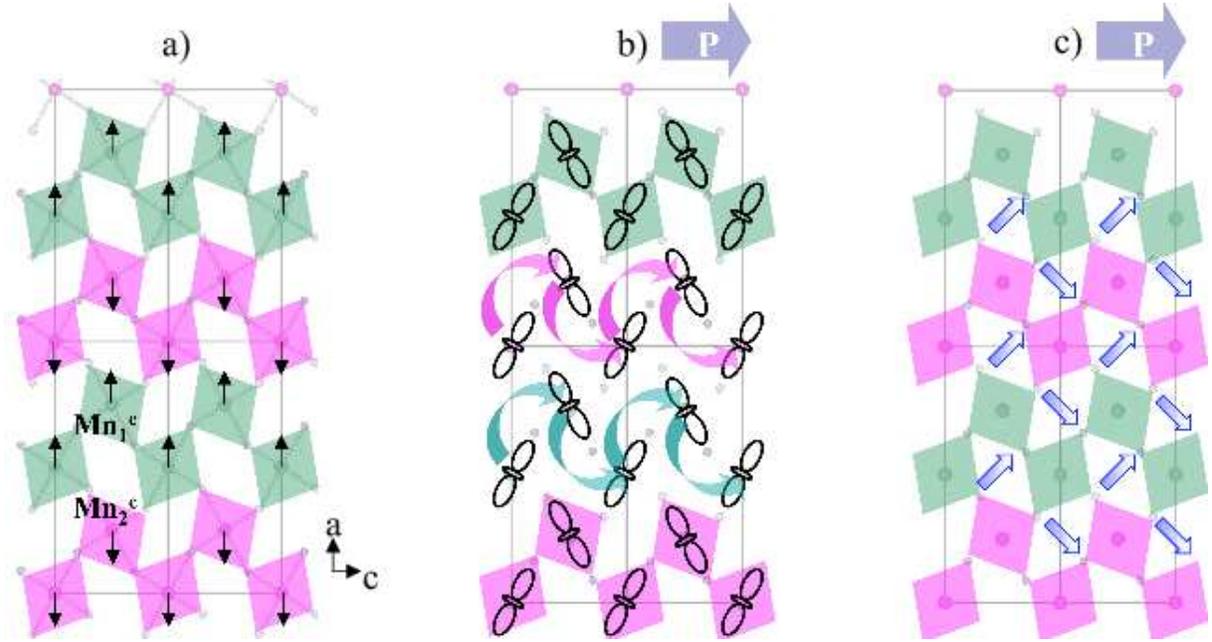}}
\caption{\label{fig:hmo} a) Ionic arrangement of AFM-E \hmo in the
  MnO$_2$ plane. Green (pink) rhombi denote in-plane projections of
  MnO$_6$ octahedra around the up-spin (down-spin) Mn ion. Spin
  directions indicated by black arrows. b) Schematic orbital-ordering
  for Mn $e_g$ states. Circular arrows show hopping paths, as induced
  by the AFM-E spin configuration; green and pink arrows denote
  asymmetric hoppings for up-spin and down-spin electrons,
  respectively. c) Schematic local dipoles (denoted by blue arrows)
  drawn from O$^{ap}$ (bonded to Mn with antiparallel spins) to
  O$^{p}$ (bonded to Mn with parallel spins). In b) and c), the
  direction of polarization is also shown.}
\end{figure}

Let us start the discussion of ferroelectricity in orthorhombic
manganites, $R$MnO$_3$, by plotting the AFM spin-arrangement
characteristic of the E-type \hmo. In Fig.~\ref{fig:hmo}a we sketch
the ions in the MnO$_2$ plane and highlight the zig-zag spin-chains,
typical features of the E-type antiferromagnetism: zig-zag
ferromagnetic (FM) spin-up-chains (green atoms in Fig.~\ref{fig:hmo}a)
are antiferromagnetically coupled to neighboring spin-down-chains
(pink atoms in Fig.~\ref{fig:hmo}a).  The out-of-plane coupling is
also AFM. We note that the antiferromagnetically-coupled zig-zag
chains lead to a doubling of the conventional GdFeO$_3$-like unit cell
(20 atoms, $Pnma$ space group) along the $a$-axis. Indeed, the E-type
was experimentally observed to be the magnetic ground state in
distorted manganites with small ionic radius for the rare-earth ion
({\em i.e.}  $R$ = Ho, \dots , Lu) \cite{expe,goodenough}. It was
shown \cite{Cheong/Mostovoy:2007,prbkyama} that the stabilization of
an $\uparrow\uparrow\downarrow\downarrow$ spin-chain (as the one
present in the E-type along the diagonal directions in the $a$-$c$
plane, {\it cfr.}  Fig.~\ref{fig:hmo}a), is driven by {\em i}) a
relatively small nearest-neighbor exchange coupling constant; {\em
  ii}) a large AFM next-nearest-neighbor; {\em iii}) a quite large
magnetic anisotropy so that the spins can be considered as Ising-like.

Why should the E-type magnetic configuration lead to a ferroelectric
polarization? This can be rationalized in different (though somewhat
inter-connected) ways, depending on the orbitals or atoms one focuses
on.

Let's start with Mn $e_g$ states. Being Mn in a $d^4$ electronic
configuration, the strong Jahn-Teller effect leads to two large and
two small in-plane Mn-O bond lengths, along with a staggered
(3$x^2-r^2$)/(3$y^2-r^2$) orbital-ordering, typical for the class of
rare-earth manganites. Within a double-exchange-like picture, this
peculiar orbital-ordering (OO) leads to a favored hopping of the
electron on the two (out of four nearest neighbors) Mn-sites towards
which the orbital is pointing. What is peculiar of the E-type (and
different from the conventional A-type in early-rare-earth manganites)
is that, out of these two Mn sites, hopping will preferentially occur
on the Mn with the spin parallel to the starting site, and not on the
other which shows an opposite spin. This ``asymmetric" hopping creates
a ``one-way path" for the electron, schematically shown by the
circular arrows in Fig.~\ref{fig:hmo}b. At this point, it is clear
that the short $c$ axis is a ``preferential" direction for the
electron, with a well-defined sign for the electron hopping. This
mechanism therefore breaks inversion symmetry and opens the way to a
ferroelectric polarization $P_c$.

Another way to explain the direction of polarization is to look at
oxygen sites. Again due to the peculiar E-type spin-configuration,
there will be two kinds of O sites: those bonded to Mn with parallel
spins (labeled as O$^p$) and those bonded to Mn with antiparallel
spins (labeled as O$^{ap}$). Due to this inequivalency, their
electronic structure will be different (even if the ions are frozen
into a centrosymmetric ``paramagnetic" configuration).  This leads to
a sort of oxygen ``charge-density wave" which can be thought of in
terms of a set of ordered dipoles resulting in a net ferroelectric
component, again only along the short $c$-axis ({\it cfr.}
Fig.~\ref{fig:hmo}c).

We would now like to make one comment related to ferroelectric
switching in IMF. As is well known, in conventional displacive
perovskite-like ferroelectrics, the switched state ({\em i.e.} the one
with $-\vec{P}$) is achieved by displacing the ions (with respect to a
reference centrosymmetric structure) in the opposite way compared to
the $+\vec{P}$ state. However, when asking how to switch $\vec{P}$ in
the case of magnetically-driven ferroelectrics, one might guess that
some changes in the spin-arrangement (rather than in the ionic
arrangement) should be involved. Indeed, from both Fig.~\ref{fig:hmo}b
and c, it is clear that $\vec{P}$ is switched by changing the
direction of half of the spins in the unit cell. For example, if we
revert the sign of the two spins in the central part of the unit cell
(labelled as Mn$_1^c$ and Mn$_2^c$ in Fig.~\ref{fig:hmo}a), then the
circular arrows in Fig.~\ref{fig:hmo}b will run in the opposite $-c$
direction; similarly, the O-related dipoles of Fig.~\ref{fig:hmo}c
will also change their sign.

So far, we have taken into account purely ``electronic" mechanisms,
occurring when considering the ions frozen into their centrosymmetric
configuration. However, it is reasonable to expect some ionic
relaxations consistent with the imposed E-type spin arrangement. For
example, according to a Heisenberg-like magnetostrictive effect, one
expects that O$^p$ will try to move so as to gain a
``double-exchange"-like energy by maximizing the Mn-O-Mn angle (recall
that the energy lowering due to double-exchange is optimal in the
ideal 180-degree case), compared to O$^{ap}$ where double-exchange is
not relevant.  These ionic relaxations break the atomic centrosymmetry
and lead to an ``ionic'' contribution to the total ferroelectric
polarization, to be added to the purely electronic one.

On the basis of this introductory background, the interpretation of
DFT results for \hmo is quite straightforward. It is however very
important to remind that, at variance with model-Hamiltonian studies
allowing the qualitative prediction of a selected phenomenon,
first-principles calculations can provide a quantitative estimate as
well. Moreover, multiferroics are very complex materials where several
competing mechanisms can occur. As such, identifying the strong and
prevailing effects can be difficult within a Hamiltonian-modelling
approach; on the other hand, all the different mechanisms are taken
into account on the same footing within DFT.

We report in Table~\ref{tab:hmo} the relevant properties calculated
within DFT, such as: {\em i}) the Mn-O-Mn angles between parallel
($\alpha^p$) and antiparallel ($\alpha^{ap}$) Mn spins, obtained after
ionic relaxations in the presence of the E-type spin arrangement; {\em
  ii}) the values of the polarization calculated in several different
ways: a purely electronic contribution ($P_{ele}^{BP}$), estimated via
the Berry-phase approach, when the ions are clamped in a
centrosymmetric $Pnma$ configuration; the polarization calculated from
the so-called "Point Charge Model" ($P_{ion}^{PCM}$), with the ions
relaxed in the ferroelectric configuration, using ``nominal" ionic
values for the charges ({\em i.e.} 3+ on Mn and Ho and 2$-$ on the O);
the total (ionic + electronic) polarization in the relaxed ionic
arrangement, calculated according to the Berry-phase approach
($P_{tot}^{BP}$); {\em iii}) the Born effective charges, {\em i.e.}
the (3,3) components of the $Z^*$ tensor for some relevant atoms:
$Z^*(\text{Mn})$, $Z^*(\text{O}^p)$ and $Z^*(\text{O}^{ap}$). We
recall that the $Z^*_{3,3}$ elements are estimated by displacing the
selected ion along the $c$ direction by a small amount (typically
about 0.01 \AA$\:$ or less) and evaluating the change in the
Berry-phase polarization along the same $c$ axis.

\begin{table}[b]
\begin{center}
\begin{tabular}{||c c || c c c || ccc  || }
\hline
        
  \multicolumn{2}{||c||}{Mn-O-Mn ($^\circ$)}&  \multicolumn{3}{|c||}{$P$ ($\mu$C/cm$^2$)}
          &  \multicolumn{3}{|c||}{$Z^*_{3,3}$ ($e^-$)}\\ 
$\alpha^p$ & $\alpha^{ap}$  & $P_{ele}^{BP}$ & $P_{ion}^{PCM}$ & $P_{tot}^{BP}$ &  $Z^*(\text{Mn})$ & $Z^*(\text{O}^p)$ & $Z^*(\text{O}^{ap})$  \\  \hline
145.3 & 141.9 & 2.1 & 3.5 & 6.1 & 3.8 & -2.6 & -3.5 \\ \hline 
\end{tabular}
\caption{Relevant calculated properties in \hmo. First two columns:
  Mn-O-Mn angles, broken down into values for the case of parallel
  ($\alpha^p$) and antiparallel ($\alpha^{ap}$) spin. Third to fifth
  columns: polarization values calculated when considering only the
  electronic polarization in the original centrosymmetric structure
  ($P_{ele}^{BP}$), or only the PCM value upon structural relaxation
  ($P_{ion}^{PCM}$) and the total Berry-phase polarization for the
  relaxed ionic coordinates ($P_{tot}^{BP}$). Sixth to eighth columns:
  (3,3) components of the Born effective charge tensors, for Mn ions
  ($Z^*(\text{Mn})$) and the two inequivalent in-plane oxygens
  ($Z^*(\text{O}^p)$ and $Z^*(\text{O}^{ap})$).}
\label{tab:hmo}
 \end{center}
\end{table}

When focusing on the Mn-O-Mn angles, we indeed note that the angle
between Mn with parallel spins is much larger than that where spins
are antiparallel, reflecting the efficiency of relaxations driven by
double-exchange mechanisms.  As for polarization, several remarks are
in order: {\em i}) one might naively expect a magnetically-induced
mechanism to be ``weak". However, this is contradicted by the purely
electronic polarization, which is noticeably large. Moreover, this is
one order of magnitude bigger than what was estimated in the case of
spin-spirals ($\le \sim $ 0.1 $\mu$C/cm$^2$): this reflects the
efficiency of the Heisenberg vs. DM term in breaking inversion
symmetry. {\em ii}) A similar consideration holds for the total
polarization. Exchange-strictive effects due to the symmetric
Heisenberg term result in ionic displacements which cooperate with the
purely electronic polarization, summing up to the appreciable value of
6 $\mu$C/cm$^2$.

So far, we have discussed the prototypical case of \hmo; however, as
previously mentioned, the E-type is the magnetic ground state for many
distorted manganites \cite{goodenough} and it is therefore interesting
to investigate how the relevant properties (with a focus on
polarization) change as a function of the rare-earth \cite{prbkyama}.
Recall that the rare-earth cation has primarily the effect of
increasing the octahedral GdFeO$_3$-like tilting as a result of
reducing the ionic size when moving, say, from La to Lu; on the other
hand, the Jahn-Teller-like distortions are weakly affected by the
rare-earth atom \cite{goodenough,prbkyama}. The structural
modifications (relative to the Mn-O-Mn angles) have in turn important
consequences on the magnetic and dipolar order. As for the former, we
have shown \cite{prbkyama} that the first-nearest-neighbor
ferromagnetic exchange-coupling constant progressively weakens upon
decreasing the ionic radius, whereas the strong
second-nearest-neighbor AFM exchange constant is more or less constant
along the series. This implies the progressive change of the magnetic
ground-state from A-type (in early rare-earth manganites) to E-type
(in late rare-earth manganites), going through the intermediate region
($R$ = Tb, Dy) where the spin-spiral occurs as ground state.  What
happens to polarization? To perform a complete investigation of the
ferroelectric properties as a function of the octahedral tilting, we
have imposed the E-type magnetic state on all the rare-earth
manganites, irrespective of the actual magnetic ground-state. This is
a typical example of a ``computer-experiment": within DFT, at variance
with real experimental samples, one can impose several different
structural, electronic or magnetic configurations (not necessarily the
ground states) to have clear insights on specific phenomena or to
separate several competing effects.

What we focus on here is the construction of Wannier functions (WF)
\cite{Marzari/Vanderbilt:1997,freimuth} for the Mn $e_g$, Mn $t_{2g}$
and O $p$ band manifolds and on the position of the WF center with
respect to the relative ionic site.  The difference between the
polarization calculated according to the point-charge-model and via
the Berry-phase approach is commonly referred to as the ``anomalous"
contribution to polarization.  As such, it reflects somewhat the
deviation from a purely ionic state or, equivalently, highlights the
covalent character of the atomic bonds and, in turn, of the electronic
structure.  Moreover, we also recall that the polarization via the
Berry-phase approach is equivalent to the sum of the displacement of
the center of each WF from the position of the corresponding ion plus
PCM contribution. The latter was shown \cite{prbkyama} to be rather
unaffected by the $R$-ion, with a value $P_{ion}^{PCM}$ $\sim$ 2 $\mu
C/cm^2$.

\begin{figure}
\begin{center}
\resizebox{80mm}{!}
{\includegraphics[angle=0]{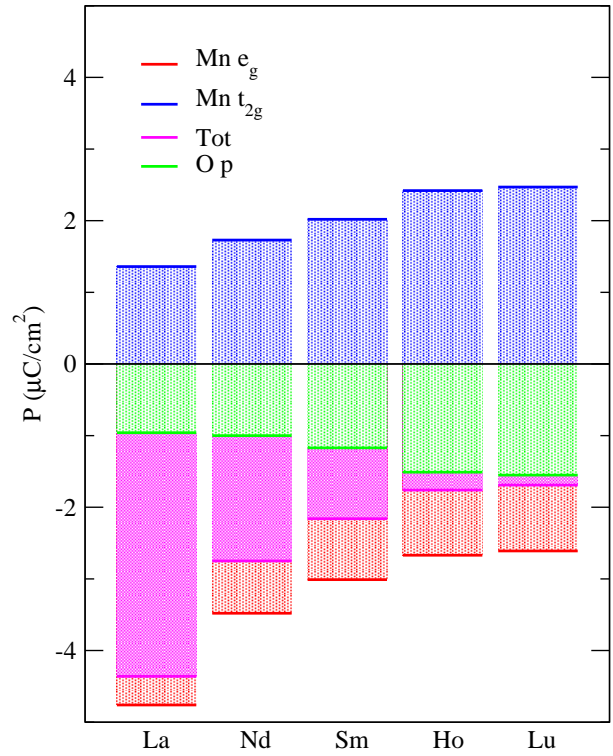}}
\end{center}
\caption{\label{fig:rmopol} Different up-spin contributions to the
  ``anomalous" term in the polarization (in $\mu$C/cm$^2$) as derived
  from WF centers: Mn $e_g$ (red), Mn $t_{2g}$ (blue), O $p$ (green)
  and total (magenta) as a function of the rare-earth ion ($R$ = La,
  Nd, Sm, Ho, Lu).}
\end{figure}

In Fig.~\ref{fig:rmopol} we report the different contributions to the
total polarization in the spin-up channel coming from the
displacements of the WF centers for the Mn $e_g$, Mn $t_{2g}$ and O
$p$, along with their sum (leading to the spin-up ``anomalous
contribution"). We note that Mn $t_{2g}$ states contribute in an
opposite way with respect to Mn $e_g$ and O $p$ states, the total $P$
having the same sign as the two latter contributions. Moreover, it is
quite clear that, whereas the O $p$ and Mn $t_{2g}$ depend relatively
little on the rare-earth ions, the $e_g$ contribution is very
sensitive to structural distortions.  Indeed, for a hypothetical
LaMnO$_3$ in the E-type spin configuration, there would be a total
polarization (coming from twice the spin-up contribution shown in
Fig.~\ref{fig:rmopol} plus the PCM term), summing up to a value
greater than 10 $\mu$C/cm$^2$! This confirms the strong sensitivity of
the $e_g$ states to the Mn-O-Mn angle: as reported in
Ref.~\cite{prbkyama}, the hopping integral strongly decreases when
moving from La to Lu, consistent with a progressively reduced band
width.  Whereas promising ways to increase P would appear in the early
rare-earth manganites (but where unfortunately the magnetic ground
state is the (paraelectric) A-type AFM), the total polarization seems
pretty much ``saturated" to a value of the order of 6 $\mu$C/cm$^2$ in
going from Ho to Lu.

We would like to comment now on the comparison with experiments. First
of all, we remark that several problems exist with the experimental
synthesis of the late $R$ manganites: indeed, the stable structure is
hexagonal, not orthorhombic \cite{synthesis,goodenough}. Modern growth
techniques, such as high-pressure high-temperature synthesis, can do
the job and synthesize ortho-manganites for late rare-earths, leading
however not to single-crystals but rather to polycrystalline
samples. This poses problems for the exact evaluation of ferroelectric
polarization, due to possible different orientations of the
polarization vector in the polycrystalline grains. To our knowledge,
there exists several values in the literature. Lorenz {\em et al.}
\cite{lorenz} reported $P \sim 0.001$~$\mu$C/cm$^2$ for \hmo, {\em
  i.e.} a value smaller by two or three orders of magnitudes than our
     {\em ab-initio} estimates. On the other hand, a much larger value
     was recently reported in AFM-E TmMnO$_3$ \cite{tmmo}: a lower
     bound of (unsaturated) polarization of about 0.15 $\mu$C/cm$^2$
     was measured, in much better agreement with our theoretical
     values. This is especially so, since Pomjakushin {\em et al.}
     \cite{tmmo} suggested that the threshold of 1 $\mu$C/cm$^2$ could
     be easily achieved in the case of single crystals. In this
     respect, we would also like to remark that the values discussed
     so far are calculated within a bare DFT approach. It is however
     well known that DFT fails in accurately modelling strong
     correlation effects, which might occur in manganites. However,
     the inclusion of an Hubbard-like correction according to the
     so-called LSDA+$U$ approach for Mn $d$ states in \hmo, lead to
     values of the polarization all larger than 1--2$\mu$C/cm$^2$ for
     $U \leq$ 8 eV.  Recently, in Ref.~\cite{maxim_electromagnons},
     the authors reported a theoretical model in the context of
     electromagnon excitations in RMnO$_3$. One of the outcome was the
     estimate of the polarization in E-type manganites based on
     optical absorption data measured for TbMnO$_3$ in the
     spiral-phase: $P$ was found to be of the order of 1
     $\mu$C/cm$^2$, therefore large and compatible with our theory
     estimates. Though some controversy is still present, there are
     more and more confirmations that the polarization in E-type is
     much higher than in the spiral phases studied so far,
     consistently with the generally accepted argument that
     magnetostrictive effects in the symmetric Heisenberg-like
     exchange should be stronger than in the antisymmetric DM part.

\subsubsection{Half-doped manganites: La$_{0.5}$Ca$_{0.5}$MnO$_3$}
\label{hdm}

Hole-doped manganites ({\em i.e.} $A_{1-x}B_{x}$MnO$_3$ where $A$ =
La, Pr, \dots and $B$ = Ca, Sr, \dots) show a rich physics, with
exciting phenomena ranging from charge-ordering to half-metallicity,
from colossal magnetoresistance to exotic phase diagrams, from
orbital-ordering to metal-insulator transitions. We will here discuss
the possibility that hole-doped manganites, with a hole-concentration
$x \sim$ 0.5, might also become ferroelectric and, therefore,
multiferroic.

\begin{figure}
\begin{center}
\resizebox{140mm}{!}
{\includegraphics[angle=90]{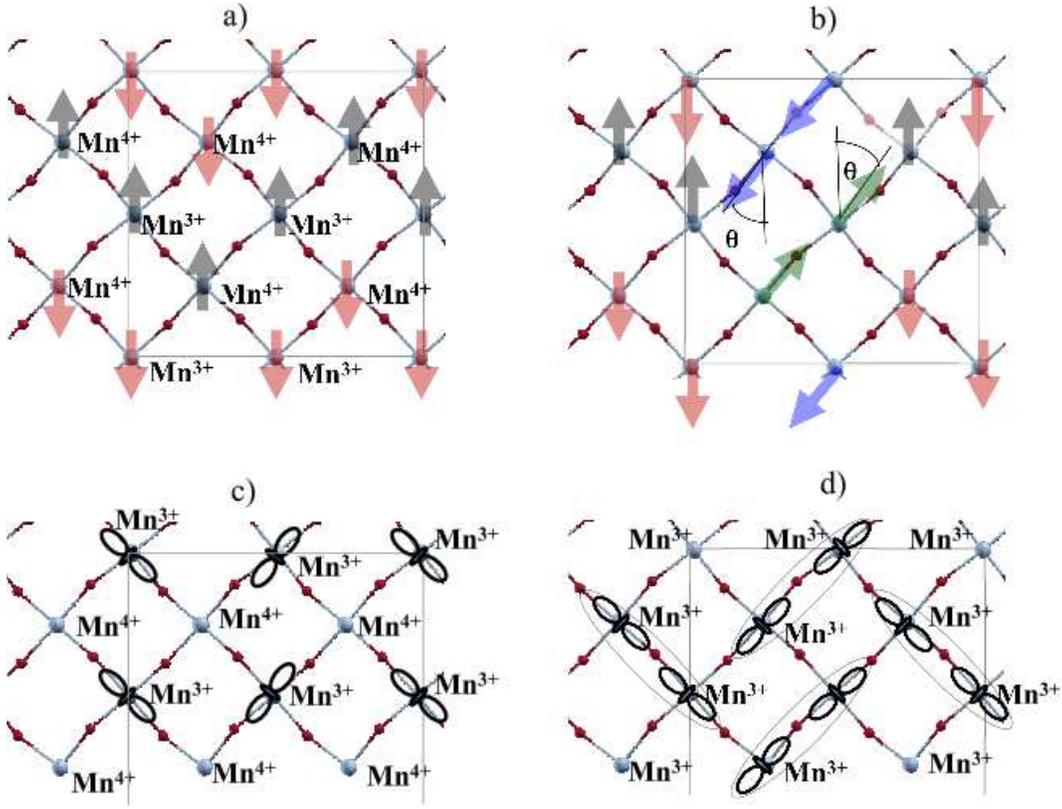}}
\end{center}
\caption{\label{fig:theta} a) Checker-board arrangement of Mn$^{3+}$
  and Mn$^{4+}$ in the MnO$_2$ plane in the SC-CO structure. The
  AFM-CE magnetic configuration is shown by double zigzag up (black
  arrows) and down (red arrows) spin chains. b) Sketch of the $\theta$
  rotation: the spins on two neighboring Mn atoms in the up-spin chain
  are rotated clockwise by $\theta$ (green arrows), along with two
  corresponding spins on neighboring Mn in the down-spin chain rotated
  clockwise by $\theta$ (blue arrows). c) The schematic
  orbital-ordering in the SC-CO structure: ideally, there should be an
  elongated Jahn-Teller-like $e_g$ orbital centered on the Mn$^{3+}$
  site and no-$e_g$-like charge on the Mn$^{4+ }$ site. d) The
  schematic OO in the BC-CO structure: the two Mn ions in the dimer
  show their $e_g$ orbitals oriented one towards each-other. ZP units
  ({\em i.e.} two Mn and the O in-between) are highlighted by
  ellipses.}
\end{figure}

La$_{0.5}$Ca$_{0.5}$MnO$_3$ (denoted in the following as LCMO) is a
very complex system from many points of view (electronic, structural,
magnetic, etc.) and, despite the many decades of work since the first
seminal paper \cite{santen}, its properties have not been clearly
elucidated. In particular, even the exact ionic coordinates and
related symmetries are still debated. Two main models have been
proposed so far: {\em a}) the first one, proposed by Radaelli {\em et
  al.} \cite{radaelli}, is based on a site-centered charge-ordered
(SC-CO) Mn$^{3+}$/Mn$^{4+}$ checkerboard arrangement in the MnO$_2$
plane (see Fig.~\ref{fig:theta}c), in which the octahedron around
Mn$^{3+}$ is Jahn-Teller-like distorted, whereas the octahedron around
Mn$^{4+}$ is rather regular; {\em b}) the second one, proposed by
Rodriguez {\em et al.} \cite{rodriguez} and referred to as a
bond-centered charge-ordered (BC-CO), is based on a structural
dimerization of Mn ions (all in a $d^4$ configuration). This leads to
a peculiar OO: at variance with the staggered OO previously mentioned
for LaMnO$_3$, here the filled Mn $e_g$ orbitals in the dimer point
one towards each other.  With respect to the mother compound,
LaMnO$_3$, there is one extra-hole every two Mn: the (spin-polarized)
hole is believed to be located on the central O in between the two
Mn. This peculiar unit (formed by two Mn and the O in between) is
often referred to as ``Zener-polaron" (ZP) \cite{daoud,wu}, after the
Zener double exchange mechanism which should be enhanced here (see
Fig.~\ref{fig:theta}d).

As far as the magnetic spin-configuration is concerned, the so-called
CE-type AFM ({\em i.e.} double zig-zag spin chains in the MnO$_2$
plane, {\it cfr.} Fig.~\ref{fig:theta}a) has been proposed as
ground-state.

We will here focus on two different mechanisms which might lead to
improper ferroelectricity in LCMO:
\begin{itemize}
\item The first one is based on breaking inversion symmetry in the
  spin-chains through a rotation (by an angle $\theta$) of the spins
  on two nearest-neighbor Mn in the up zigzag chain, along with a
  corresponding rotation of two spins in the down spin-chain ({\it
    cfr.}  Fig.~\ref{fig:theta}b), so as to keep a global AFM
  character.  This follows the theoretical proposal put forward by
  Efremov {\em et al.}  \cite{efremov}, who first suggested the
  possibility of multiferroicity in manganites. According to
  Ref.~\cite{efremov}, such rotation should progressively lead from a
  fully SC-CO (in the ideal CE-type, $\theta$ = 0$^\circ$) to a fully
  BC-CO for $\theta$ = 90$^\circ$ (where the dimerization process
  driven by spin ordering is maximized). Efremov {\em et al.}
  predicted that, in both the extreme cases, $\theta$ = 0$^\circ$ and
  $\theta$ = 90$^\circ$, the polarization should vanish: for $\theta$
  = 0$^\circ$, the checkerboard arrangement should be fully
  centrosymmetric (both structurally and electronically), whereas for
  $\theta$ = 90$^\circ$ the Mn should not show any
  charge-disproportionation. However, for in-between values of
  $\theta$, the intermediate SC-CO/BC-CO should lead to a small
  charge-disproportionation and, therefore, to inequivalent Mn (at
  variance with the ZP state and reminiscent of the site-centered
  CE-type). In this case, inversion symmetry would be broken by
  spin-dimerization, therefore paving the way to ferroelectricity;
\item The second mechanism occurs in the structure experimentally
  proposed by Rodriguez {\em et al.} \cite{rodriguez}. The related
  unit cell shows a ``{\em structural}" Mn-Mn dimerization and implies
  a realization of a BC-CO, not invoking (non-collinear) magnetic
  mechanisms as in the previous case, but rather thanks to electronic
  rearrangement --- such as OO --- following the structural
  distortions. Still, in this case, our mechanism for multiferroicity
  is once more a (collinear) magnetically induced mechanism based on
  the inequivalency of some specific oxygen atoms, as will be detailed
  below.
\end{itemize}

Due to the large unit-cell (80 atoms, needed to simulate the CE-type
AFM ordering, along with a checkerboard arrangement of La and Ca
cations) and the need of non-collinear spin magnetism (needed to
simulate finite values of $\theta$), the computational cost of these
simulations is very high. For this reason, the ionic positions were
not optimized within DFT, but were rather kept frozen in the structure
proposed either by Radaelli \cite{radaelli} or by Rodriguez
\cite{rodriguez}, labelled in what follows as LT-M or by LT-O,
respectively. Unfortunately, the lack of ionic minimization forbids
any DFT prediction of the actual structural and magnetic ground-state
from total-energy arguments; this calls for future studies. From our
calculated values for unrelaxed structures, it seems that the CE with
SC-CO is the phase showing lowest total energy; however, for example,
the SC-CO state with a rotation $\theta$ = 45$^\circ$, is higher in
energy by only $\sim$ 4 meV/Mn. One can therefore conjecture that, in
real samples, there might be a coexistence of nanoscale regions with
different magnetic structures ({\em i.e.} with zero and finite
$\theta$ values).

\begin{figure}
\begin{center}
\resizebox{160mm}{!}
{\includegraphics[angle=90]{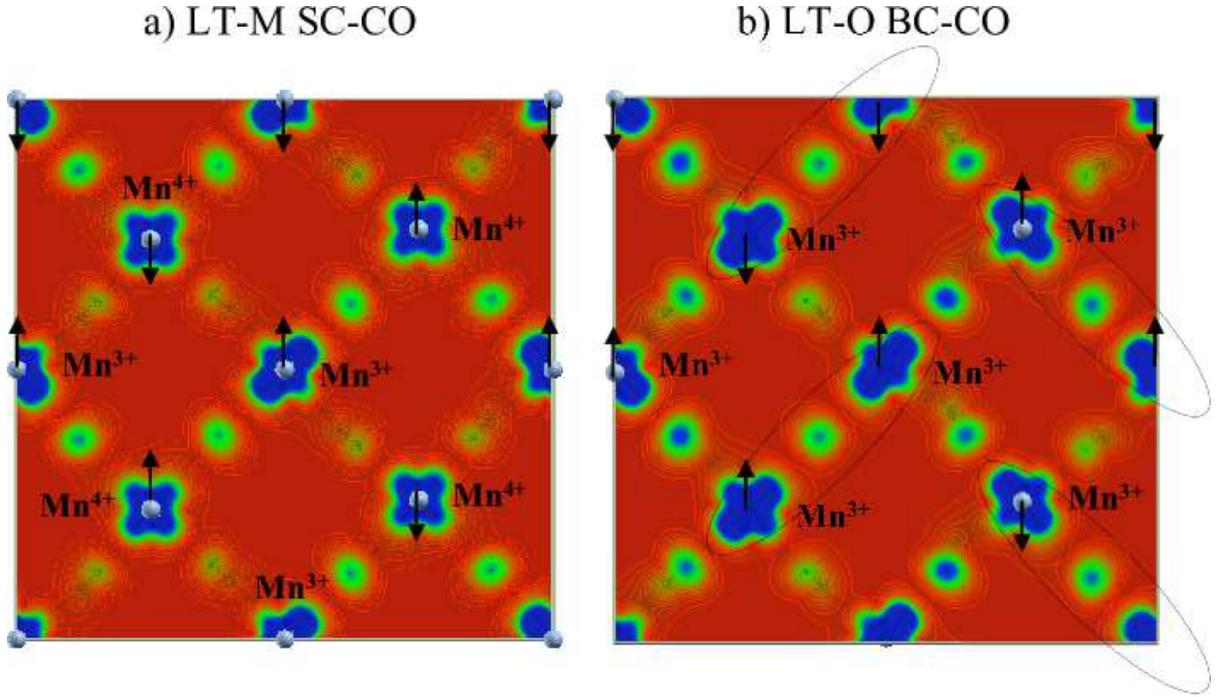}}
\end{center}
\caption{\label{fig:parchg} Isolines of the $e_g$ charge in a) the
  LT-M SC-CO and b) the LT-O BC-CO structures. Red (blue) lines marks
  the minimum (maximum) charge, through the intermediate green
  lines. In a), black arrows mark the Mn spin directions. In b), ZP
  are highlighted.}
\end{figure}

Before discussing the relevant ferroelectric properties, let us
mention some general features in the electronic structures of the LT-M
and LT-O systems, both in the CE-type AFM spin configuration ({\em
  i.e.} $\theta$ = 0). In Fig.~\ref{fig:parchg} we show the isolines
of the electronic charge plotted in the energy region where the Mn
$e_g$ states are located. It is clear that in the LT-M
(Fig.~\ref{fig:parchg}a) the shape of the $e_g$ electronic cloud,
centered on the ``nominal" Mn$^{3+}$, is markedly elongated towards
the neighboring Mn$^{4+}$ with parallel spins. On the other hand, the
Mn$^{4+}$ show a very isotropic distribution of the charge. The
situation is different in the LT-O structure (Fig.~\ref{fig:parchg}b),
where the OO clearly shows the $e_g$ orbitals forming ``dimers" with
their charge distribution pointing one towards the other, as driven by
the underlying ionic configuration.  Let us mention a note on the CO:
consistently with previous reports, the actual
charge-disproportionation in LCMO within DFT is of the order of only
0.1--0.2 electrons in the LT-M SC-CO, at variance with the ideal
situation of ``full" charge disproportionation, where the $e_g$
electron cloud should be completely distributed around the Mn$^{3+}$,
with no-charge on the Mn$^{4+}$. In this sense, the calculated OO in
the LT-M ({\it cfr.} Fig.~\ref{fig:parchg}a) is different from the
nominal situation ({\it cfr.} Fig.~\ref{fig:theta}c) with clear
signatures of $e_g$ charge also around the Mn$^{4+}$. We remark,
however, that the small charge-disproportionation detected in the LT-M
structure becomes really negligible ($<$0.02 electrons) in the LT-O
BC-CO; this suggests that it is still meaningful to consider the LT-M
$\rightarrow$ LT-O transition as a corresponding SC-CO $\rightarrow$
BC-CO transition.
 
Let's now consider what happens in the LT-M structure upon increasing
$\theta$ from the initial zero-value: our calculated electronic
structures (not shown, see Ref.~\cite{gglcmo}) indicate a decreasing
$e_g$ band-width and a related increasing band-gap. This can be
rationalized by comparing the spin-arrangement with finite $\theta$
with the original CE-type.  Upon spin rotation, the $e_g$ electron ---
which could hop equivalently on the two nearest-neighbor Mn$^{4+}$ on
both sides along the spin-chain in the CE-AFM phase --- will now
preferentially hop on the Mn$^{4+}$ which shows a parallel spin, since
hopping in the other direction is prevented by the spin
misalignment. This effect would rather lead to a decreased hopping and
to a reduced $e_g$ band-width, at variance with our findings. However,
one needs to consider that, for $\theta \neq$ 0, there will be an
increasing probability to hop on the neighboring spin-chain (prevented
by opposite spin configuration in the CE-AFM phase). Overall, there
will be therefore an increased hopping integral, consistently with our
findings and with Hamiltonian-modelling studies, as well
\cite{gglcmo}.

For our purposes, the most important finding is that a finite $\theta$
in the LT-M induces a rather large polarization, as shown in
Table~\ref{tab:P} (first line), with an increasing parabolic trend of
$P$ vs. $\theta$. We note that this is a purely electronic
polarization, since the ions are fixed in their centrosymmetric
arrangement \cite{radaelli}. The Heisenberg-like symmetry breaking ---
as driven by spin-rotation --- is therefore confirmed as an efficient
tool to induce large ferroelectricity (recall that spin-orbit coupling
and the related DM interaction is neglected in the present context).

\begin{table}[b]
\begin{center}
\begin{tabular}{|c| c| c | c | c | c| }
\hline \hline
& 0$^\circ$ & 22.5$^\circ$ & 45$^\circ$ & 67.5$^\circ$ & 90$^\circ$ \\ \hline \hline
 LT-M     & 0.0  & 0.19 & 0.66 & 1.56& 2.70 \\ \hline
 LT-O    &  7.18& 6.62  & 5.13  & 2.84  &0.0  \\ \hline   \hline
\end{tabular} \end{center}
\caption{Berry-phase polarization (in $\mu$C/cm$^2$) calculated in the
  LT-M and LT-O structure as a function of spin-rotation angles
  $\theta$ (first line).}
\label{tab:P}
\end{table}

\begin{figure}
\begin{center}
\resizebox{100mm}{!}
{\includegraphics[angle=90]{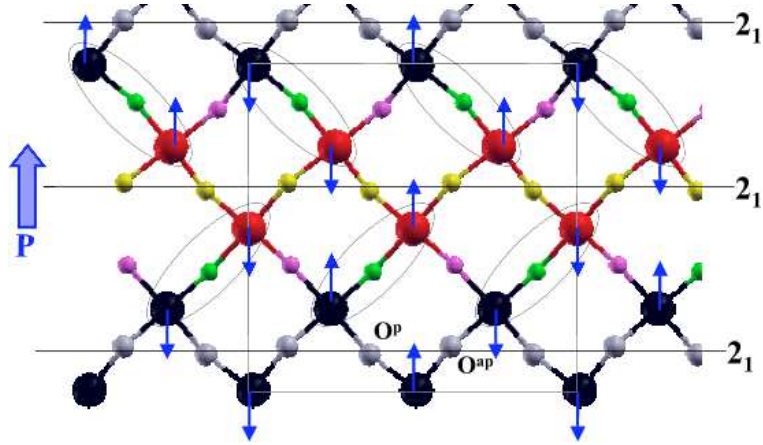}}
\end{center}
\caption{\label{fig:p21} Atomic configuration in the MnO$_2$ plane of
  the LT-O structure: symmetry-equivalent atoms are marked in the same
  color. Note that red and black spheres mark Mn atoms: despite being
  symmetry-inequivalent, the two kinds of Mn are only marginally
  different from the electronic point of view, with small differences
  in the Mn-O bond-lengths (see Ref.~\cite{rodriguez}.). Horizontal
  lines mark the two $2_1$ screw axes in the unit cell, crossing the O
  atoms (marked as grey and yellow). The blue arrows on the Mn ions
  denote the spin directions in the AFM-CE spin configuration: when
  considering the spin-directions, the grey atoms (structurally
  equivalent by symmetry) become electronically different: they are
  alternatively bonded to two parallel Mn spins and to two
  anti-parallel spins (see labels on two selected oxygens).}
\end{figure}

We will now focus on the calculated values of $P$ in the LT-O
structure (see Table~\ref{tab:P}) and start the discussion for the
$\theta$ = 0 case. Without any magnetic ordering imposed and as
determined experimentally, the LT-O structure shows a $P2_1nm$ space
group: this implies that some in-plane oxygens are structurally
equivalent (shown in the same color in Fig.~\ref{fig:p21}), due to the
$2_1$ screw symmetry. However, when imposing the AFM-CE
spin-configuration, the Oxygen equivalency is lifted: there is an
alternation of O$^p$ bonded to two parallel spins and of $O^{ap}$
bonded to two antiparallel spins. This is sufficient to give rise to
ferroelectricity in the direction shown in
Fig.~\ref{fig:p21}. Remarkably, the induced polarization reaches the
surprisingly large value of several $\mu C/cm^2$. To further verify
that a magnetically-induced mechanism is the source of the
ferroelectricity, we have also performed a $\theta$-like rotation of
the spin dimers, similar to the previous case of the LT-M ({\it cfr.}
Fig.~\ref{fig:theta} b)). In this case, upon spin-rotation, the
inequivalency of the Oxygens crossing the $2_1$ axis is reduced. In
the extreme situation, $\theta$ = 90$^\circ$, all the O atoms are now
bonded to two Mn with perpendicular spins: in this configuration, they
all look equivalent and the source of polarization vanishes. Indeed,
DFT calculations confirm that this is the case ({\it cfr.}
Table~\ref{tab:P}).  In summary, our DFT results (both from \hmo and
LCMO) offer a confirmation that the O inequivalency is an efficient
handle to achieve and/or tune a large ferroelectric response.

\subsection{Problems and perspectives in Improper multiferroics}
\label{perspectives}

As indicated by the huge interest in the last few years,
magnetically-driven ferroelectrics, with ortho-TbMnO$_3$ taken as
prototype, are with no doubt an exciting class of materials. However,
there are a few bottlenecks which prevent their use in large-scale
applications: {\em i}) their polarization is generally very small
($\leq$ 0.1 $\mu$C/cm$^2$); {\em ii}) their ordering temperature is
very low (of the order of few tens of K); {\em iii}) being globally
antiferromagnets, their net magnetization is always zero (a
ferromagnetic spin ordering alone cannot break inversion symmetry!).
In this respect, we will certainly see some activity in future years
to get rid of these problems.

As shown in this review, at least point {\em i}) can be beautifully
overcome when considering Heisenberg-like exchange-striction, as shown
in E-type manganites. The ordering temperature of the latter is,
however, extremely low ($T_N$(\hmo) $\sim$ 26 K). One possibility to
increase the ordering temperature without losing the
non-centrosymmetric Heisenberg-like exchange-striction is to consider
rare-earth nickelates \cite{ggni} (for example, $T_N$(HoNiO$_3$) = 145
K, $T_N$(LuNiO$_3$) = 130 K, etc.). Nickelates are rather complex
materials, with several important issues still under debate, including
the origin of their metal-insulator transition as well as their spin
configuration. As for the latter, both non-collinear and collinear
spin-arrangements have been put forward from neutron diffraction
studies \cite{scagnoli,alonso}. In addition, nickelates show a
charge-disproportionation: Ni ions, in the nominal 3+ valence-state,
split into two groups of Ni$^{2+}$ and Ni$^{4+}$ \cite{cd}. This adds
one degree of freedom to achieve ferroelectricity. For example, as
suggested in Ref.~\cite{jvdbjpcm}, one of the proposed magnetic
configurations shows, along the [111] direction, a sequence of
Ni$^{2+}$-Ni$^{4+}$-Ni$^{2+}$-Ni$^{4+}$ as for charge-ordering and a
sequence of $\uparrow\uparrow\downarrow\downarrow$ planes as for
spin-ordering. The combination of spin and charge-ordering would break
centrosymmetry, leading to a polarization along the [111] direction.
Another spin-configuration, proposed by experiments, seems to be very
similar to the E-type in \hmo, the only difference being the stacking
of TMO$_2$ (TM = Mn, Ni) planes: whereas the out-of-plane coupling is
always AFM in \hmo, in nickelates there are NiO$_2$ alternatively
coupled ferromagnetically and antiferromagnetically. However, the
different out-of-plane stacking does not destroy the mechanism for
polarization, induced in a way very similar to \hmo.  Our preliminary
calculations \cite{ggni} show that the two mentioned collinear
magnetic ground-states in monoclinic $R$NiO$_3$ ($R$ = Ho, Lu) are
basically degenerate ({\em i.e.} the differences in total energies are
below our numerical uncertainty). Consistently with a
Heisenberg-driven mechanism, both spin-configurations give rise to a
large polarization (of the order of few $\mu$C/cm$^2$) along different
directions, suggesting nickelates as a new and interesting class of
magnetically-driven multiferroics.

Going back to the bottlenecks mentioned above, point {\em iii}) might
be overcome by considering magnetite.  In this review, we have
discussed so far a few examples where spin-ordering is a necessary
ingredient to break inversion symmetry. However, there are materials
in which the polarization is induced purely by charge-ordering, such
as LuFe$_2$O$_4$ and Fe$_3$O$_4$ below the Verwey transition
temperature ({\em i.e.}  corresponding to the metal-insulator
transition, $T_V \sim$ 120 K). In magnetite, the spin-arrangement is
ferrimagnetic ({\em i.e.} tetrahedral and octahedral Fe sites show up
and down spin, respectively). The role of magnetism, however, does not
seem to be relevant for polarization. Magnetite is a complex and
controversial system: the Fe$^{2+}$/Fe$^{3+}$ charge ordering pattern
on octahedral iron sites is still under debate \cite{wright,rixs}.
However, the $Cc$ symmetry has been proposed by diffraction studies
and confirmed from first-principles to be the ground state
\cite{jeng}.  In the $Cc$ case, octahedral Fe sites, form a
corner-sharing tetrahedron network: 75$\%$ of the tetrahedra show the
so-called "3:1" pattern (meaning that, in each tetrahedron, 3 sites
are Fe$^{2+}$ and one is Fe$^{3+}$ or vice versa), whereas 25$\%$ show
a 2:2 pattern (meaning that 2 sites are Fe$^{2+}$ and two are
Fe$^{3+}$ in the tetrahedron). It happens that the $Cc$ is
non-centrosymmetric; indeed, our DFT calculations \cite{alexe} show
the polarization induced by charge-ordering to be of the order of few
$\mu$C/cm$^2$, suggesting magnetite to be the first improper
multiferroic known to mankind.

\section{Summary and Conclusions}

In summary, we have presented some examples which show the power of
DFT-based methods in the field of multiferroic materials. This
includes: {\em i}) rationalizing experimental observations in known
multiferroics, {\em ii}) designing new (artificial) multiferroics with
optimized properties (larger ferroelectric polarization, strong
ferromagnetism, higher ordering temperatures, etc.), and {\em iii})
proposing and quantifying novel microscopic mechanisms, based on
electronic degrees of freedom, which potentially lead to
ferroelectricity in magnetic transition metal oxides.

It is apparent that the field of proper magnetic ferroelectrics has a
relatively long history: many of these materials have already been
studied in the 1960s or later, but have only recently been
rediscovered. Due to substantial advancements in experimental
synthesis and characterization techniques on one side, and the
availability of powerful computational methods together with new
theoretical approaches on the other side, substantial progress in
understanding these materials has been achieved during recent
years. Similar to the the case of non-magnetic ferroelectrics,
first-principles calculations have shown a remarkably high degree of
accuracy, reliability, and predictive capability for the class of
proper multiferroics. Nevertheless, many open questions still remain,
in particular how to achieve large polarization, large magnetization,
and strong magneto-electric coupling above room temperature, or what
mechanisms for coupling between magnetic and ferroelectric properties
do exist in these materials.

On the other hand, the field of DFT calculations for improper
multiferroics is only a couple of years old. As such, it is not clear
at the moment how accurate the predictive capabilities of current DFT
approaches are for relevant quantities such as structural or
electronic properties and, most importantly, polarization. On the
experimental side, the synthesis of some compounds ({\em i.e.} as
shown for ortho-manganites with late rare-earth ions) is not under
full control, making the theory-experiment comparison rather
complicated. On the modelling side, the role of electronic
correlations (where DFT often shows its limits) is certainly more
relevant in improper than in proper magnetic ferroelectrics.  In this
respect, future developments on the theory side ({\em i.e.} invoking
novel exchange-correlation functionals to better describe many-body
effects) are desirable. As such, a strong interaction with the
experimental and model-Hamiltonian communities active in the field, as
well as the extension of DFT studies to a much larger set of materials
(showing different microscopic mechanisms or simply different
chemical, structural, or electronic properties), will be necessary to
achieve a satisfactory qualitative and quantitative description of the
complex physics at play in improper multiferroics.

\bigskip

{\bf \large Acknowledgements}

Part of the research leading to the presented results has received
funding from the European Research Council under the EU Seventh
Framework Programme (FP7/2007-2013)/ERC grant agreement
n. 203523. C.E. acknowledges support by Science Foundation Ireland
under Ref. SFI-07/YI2/I1050.

\end{document}